\documentclass[journal]{IEEEtran}
\usepackage{ifpdf}

\usepackage{cite}

\usepackage{multicol}
\usepackage{graphicx}
\usepackage{epstopdf} 
\usepackage{xcolor}
\usepackage{tensor}
\ifCLASSINFOpdf

\else

\fi

\usepackage[cmex10]{amsmath}
\usepackage{amsfonts,amssymb}

\usepackage{algorithmic}

\usepackage{array}

\ifCLASSOPTIONcompsoc
\usepackage[caption=false,font=normalsize,labelfont=sf,textfont=sf]{subfig}
\else
\usepackage[caption=false,font=footnotesize]{subfig}
\fi

\usepackage{stfloats}
\usepackage{url}

\usepackage{arydshln}
\usepackage{enumerate}
\usepackage{mathtools}

\usepackage[mathcal]{euscript}
\newtheorem{theorem}{Theorem}
\newtheorem{lemma}{Lemma}

\newtheorem{defn}{Definition}

\newtheorem{remark}{\textbf{Remark}}
\newcommand{\vect}[1]{\mathbf{#1}}
\newcommand{\y}[1]{\mathbf{#1}}

\newcommand{\bs}[1]{\boldsymbol{#1}}
\def\beq{\begin{equation}}
\def\eeq{\end{equation}}

\DeclarePairedDelimiter\ceil{\lceil}{\rceil}
\DeclarePairedDelimiter\floor{\lfloor}{\rfloor}

\graphicspath {{Figures/}}
\IEEEoverridecommandlockouts

\begin{document}

	\title{Joint Transmit and Circuit Power Minimization in Massive MIMO with Downlink SINR Constraints: When to Turn on Massive MIMO?}

	\author{Kamil~Senel,~\IEEEmembership{Member,~IEEE,}
	        Emil~Bj\"{o}rnson,~\IEEEmembership{Member,~IEEE,}
	        and~Erik~G.~Larsson~\IEEEmembership{Fellow,~IEEE}
	        \thanks{\copyright 2019 IEEE. Personal use of this material is permitted. Permission from IEEE must be obtained for all other uses, in any current or future media, including reprinting/republishing this material for advertising or promotional purposes, creating new collective works, for resale or redistribution to servers or lists, or reuse of any copyrighted component of this work in other works.}
	        \thanks{Parts of this work will be presented at the IEEE Wireless Communications and Networking Conference (WCNC), 2018 \cite{kamil2017wcnc}.}
	\thanks{The authors are with the Department of Electrical Engineering, (ISY), Link\"{o}ping University, Sweden.}
	\thanks{This work was supported by ELLIIT and the Swedish Foundation for Strategic Research (SSF).}}

	\maketitle

	\begin{abstract}
In this work, we consider the downlink of a multi-cell multiple-input multiple-output
(MIMO) system and find the jointly optimal number of base station (BS) antennas and transmission powers that minimize the power consumption while satisfying each user's effective signal-to-interference-and-noise-ratio (SINR) constraint and the BSs' power constraints. Different from prior work, we consider a power consumption model that takes both transmitted and hardware-consumed power into account. We formulate the joint optimization problem for both single-cell and multi-cell systems. Closed-form expressions for the optimal number of BS antennas and transmission powers are derived for the single-cell case. The analysis for the multi-cell case reveals that increasing the number of BS antennas in any cell always improves the performance of the overall system in terms of both feasibility and total radiated power. A key contribution of this work is to show that the joint optimization problem can be relaxed as a geometric programming problem that can be solved efficiently. The solution can be utilized in practice to turn on and off antennas depending on the traffic load variations. Substantial power savings are demonstrated by simulation.           
	\end{abstract}


	\IEEEpeerreviewmaketitle

	\section{Introduction}\label{sec:Introduction}
	Massive multiple-input multiple-output (MaMIMO) 
	is a physical layer technology in which the BSs are equipped with a large number of antennas \cite{redBook}. This allows MaMIMO systems to spatially multiplex tens of users on the same time-frequency resources, which greatly enhances the spectral efficiency of the network \cite{erik2017Tech}. Furthermore, simple signal processing techniques, such as linear precoding and detection schemes, are close to optimal due to the quasi-orthogonality of the channels \cite{emil10myths}. 	
	 These features makes MaMIMO a key technology for future wireless networks \cite{andrews2014what5g}. 
 
 A crucial design criterion for $5$G systems is improved energy efficiency \cite{prasad2017EEinMIMO}. For MaMIMO systems, it has been shown that the required downlink transmission power to maintain a constant SINR is inversely proportional to the number of BS antennas \cite{hoydis2013howManyAntennas}. However, the total power consumption does not necessarily decrease with increased number of antennas when using a realistic power consumption model that also takes the hardware-consumed power into account \cite{desset2016eeMIMO,desset2014PowerConsumption}.    
   
  There are various studies on the energy efficiency of MaMIMO systems, which is commonly defined as the ratio between data rate and the power consumption. Numerical analysis has shown that MaMIMO systems have the potential to improve the energy efficiency up to a factor of $1000$ compared to a setup with typical LTE BSs \cite{yang2013EEgreen}. For a given uplink sum rate, the optimal numbers of BS antennas and users are investigated for a single-cell system in \cite{mohommed2014ZFee}. However, the cost of acquiring channel state information (CSI) is ignored, which can lead to misleading results. Another work that considers the energy efficiency problem with perfect CSI at the receivers and provides an iterative approach is \cite{ng2012ee}. In \cite{emil2015IsMIMOanswer}, the joint uplink-downlink energy efficiency is maximized with respect to the number of BS antennas, data rate, number of users, and transmission powers. In a multi-cell setup, the energy efficiency optimization problem with respect to the number of BS antennas is investigated in \cite{emil2015eeMaMIMO}, which assumes perfect CSI. The total power consumption for a single-cell system is minimized in \cite{victor2015MIMOatNight} under a low traffic assumption.    
  These prior works can provide general insights into the deployment of energy-efficient networks, but when the network is in place, one cannot optimize the data rates or number of users---these are given by the actual user traffic.
  
  	When the traffic load is low, the energy efficiency of a multiantenna systems can be increased by utilizing only a subset of the available BS antennas, referred to as antenna selection. Such an approach allows the BS to utilize only the antennas that, for the current small-scale fading realization, provide a high contribution to the SNR. An overview of MIMO systems with antenna selection is presented in \cite{andy2004atas}. Antenna selection based on small-scale fading realization was recently studied in the context of MaMIMO \cite{arash2017atas,lee2017enhancedATAS}. It has been shown that antenna selection provides a significant increase in capacity when the number of antennas is greater than RF chains and perfect CSI is available \cite{gao2013convexOptAtas}. A genetic algorithm for antenna selection that is capable of optimizing different objective functions is proposed in \cite{makki2017geneticATAS}. In \cite{benmimoune2017Atas}, a subset of the antennas is selected in a decentralized manner by the receiving users under a setup with imperfect CSI. An iterative water-filling scheme for antenna selection is presented in \cite{xu2013EEinMIMOBC}. A detailed comparison of antenna selection approaches is presented in \cite{le2016TAS}. 
  	In principle, the unselected antennas can be turned off to save energy, but in wideband systems with many subcarriers it is unlikely that a certain antenna is simultaneously unselected on all subcarriers; thus, antennas can generally not be turned off. Furthermore, antenna selection based on small-scale fading realizations requires all antennas to be turned on during the channel estimation, thus the sleep time is very short.
  	
In this work, we consider the downlink of a MaMIMO system and minimize the total power consumption by jointly optimizing the transmission powers and the number of active antennas, with given constraints on the maximum transmit power, the required effective SINRs, and available number of antennas. The optimization is based on the large-scale fading coefficients, not the small-scale fading realizations, which enables us to turn off antennas to save power when the traffic load is low. For downlink precoding, maximum ratio transmission (MRT) and zero-forcing (ZF) are considered. A conference version of this work has been presented at \cite{kamil2017wcnc}, where only a single-cell setup is considered. This paper contains more general and complete results for the single-cell case along with the investigation of multi-cell setups. To summarize, the main contributions of this work are as follows:
 
  		\begin{itemize}
  			\item The number of BS antennas and transmission powers are jointly optimized, while satisfying individual power and SINR constraints for both single-cell and multi-cell setups. The feasibility conditions are manifested analytically. The problem can be infeasible for a given number of antennas, but always becomes feasible as the number of antennas increases for a single-cell system. (Lemmas \ref{lem:SC-MmaxFeasibility}-\ref{lem:SC-MexpandsFeasibility set})        	
  			\item We provide closed-form expressions for the optimum number of antennas and transmission powers when using MRT or ZF precoding for a single-cell system. We compare these schemes analytically and numerically. (Theorem \ref{thm:thm-2-SC-optM}, Lemma \ref{lem:SC-optMdiffMRC-ZF}) 		 			 
  			\item We prove that increasing the number of BS antennas in any cell does not deteriorate the performance of the overall system. (Lemma \ref{lem:MC-diffMdiffP}) 
  			\item For the multi-cell case, we reformulate the joint optimization problem as a geometric programming problem that can be solved efficiently.       			 
  		\end{itemize}

\subsection{Organization and Notation}
	
  Vectors and matrices are denoted by boldface lowercase and uppercase letters, respectively. The superscripts $(\cdot)^T$ and $(\cdot)^H$ represent the transpose and conjugate transpose. $\vect{I}_N$ is the $N\times N$ identity matrix. The inverse of a matrix $\vect{A}$ is denoted by $\vect{A}^{-1}$ and $r(\vect{A})$ is its spectral radius.
   $\mathbb{R}, \mathbb{Z}$ are the sets of real numbers and integers whereas the strictly positive integers are represented by $\mathbb{Z}^+$. The trace operator is denoted by $\text{tr}(\cdot)$ and $\|\cdot\|$ is used for the Euclidean norm. The $(i,j)$-th element of a matrix $\vect{A}$ is denoted by $a_{ij}$ and similarly the $m$th element of a vector is described by $[\vect{a}]_m$. For matrices/vectors $\succeq, \succ$ operators are used for component-wise inequalities. 
  
 The rest of the paper is organized as follows. The system setup is introduced in Section \ref{sec:SystemSetup}. The single-cell setup is investigated in Section \ref{sec:SC system}, including the closed-form solution for the joint optimization problem. In Section \ref{sec:MC system}, the results are extended to multi-cell systems and the joint optimization problem is reformulated as a geometric programming problem. The numerical analysis is presented in Section \ref{sec:Numerical} and Section \ref{sec:Conclusion} concludes the paper.  
  
	\section{System Setup}\label{sec:SystemSetup}

	We consider a massive MIMO system with $L$ cells, indexed by $l$, and each cell has a BS with $M_l$ active antennas, out of $M_{\max}$ available antennas. Cell $l$ contains $K$ single-antenna users that simultaneously communicate with the BS. 
	The system operates in time-division duplex (TDD) mode, where time-frequency resources are divided into coherence intervals, such that each channel is constant and frequency-flat in each interval. The channels are assumed to take independent realizations in each coherence interval from stationary ergodic processes. A coherence interval consists of three phases: uplink data transmission, uplink training, and downlink data transmission.   
	In this paper, we focus on the uplink training and downlink data transmission, while we leave uplink data transmission analysis as future work.

		Let $N$ denote the length of the coherence interval (in samples) and assume that $N_p < N$ samples are used for uplink training. The remaining $N - N_p$ samples are used for downlink data transmission. During the training, the users simultaneously transmit their pilot sequences, which are known to BS and the channels are estimated based on the received signals. As in practice, the active	user set and the corresponding SINR constraints in a given coherence interval are assumed to be determined by a preceding MAC-layer process. 
			
	\section{Single Cell System} \label{sec:SC system}
	 We first analyze a single-cell system and extend the results to  multi-cell systems in Section \ref{sec:MC system}. Since we consider a single cell, we drop the cell index in this section. 
	 In the training phase, the users concurrently transmit $N_p$-length pilot sequences. The pilot sequences are assumed to be orthogonal and therefore the length must be greater than the number of users, i.e., $N_p \geq K$. The channel between user~$k$ and the BS is represented by 	 	
	\begin{equation}\label{eq:SC-channel}
	\vect{g}_k = \sqrt{\beta_k} \vect{h}_k
	\end{equation} 
    where $\beta_k$ denotes the large-scale fading and $\vect{h}_k$ is the $M \times 1$ vector representing the small-scale fading. The elements of $\vect{h}_k$ are assumed to be i.i.d. $\mathsf{CN}(0,1)$. The large-scale fading is assumed to be known at BS whereas the small-scale fading is to be estimated in each coherence interval\footnote{Real measurements performed with both uniform linear array and uniform cylindrical array antennas reveal that correlation between antennas yields only a minor penalty on the maximum rate of the users compared to the i.i.d. Rayleigh fading case \cite{gao2015MaMimoMeasurements},\cite{fredrik2013ScalingMimo}.}. The minimum mean square estimator (MMSE) 
    is utilized to obtain an estimate of $\vect{g}_k$ \cite{Kay1993estimation}.  
        Since the channels are statistically identical across all antennas, the mean square of the channel estimate is same for all $m \in \{1, \ldots,M\}$ and for the $m$th component it is given by 
    \begin{eqnarray}
    \gamma_k
    &=& \frac{N_p\rho_{ul}\beta_k^2}{1 + N_p\rho_{ul}\beta_k}. 
    \end{eqnarray}
    	where $\rho_{ul}$ denotes the uplink SNR.
    	
   In the downlink, the BS precodes and scales the data symbols to generate the transmit signal.   
    	Throughout this work, we only consider zero-forcing and maximum-ratio transmission. Although there are other methods capable of achieving somewhat better SINRs \cite{emil2014OptBF}, closed-form expressions of the effective SINRs for these methods are only available in the asymptotic region. Furthermore, ZF and MRT are nearly optimal when $M\gg K$ under high and low SINR conditions, respectively \cite{redBook}.
        Let $p_k$ be the normalized transmission power for user $k$, then the effective SINR is given by
    \begin{equation}\label{eq:SC-SINRforMRC}
    \text{SINR}_k^\mathrm{MRT} = \frac{M \gamma_k p_k}{1 + \beta_k \sum_{k' = 1}^K p_{k'}}   
    \end{equation}
    for MRT and 
        \begin{equation}\label{eq:SC-SINRforZF}
    \text{SINR}_k^\mathrm{ZF} = \frac{\left(M-K\right) \gamma_k p_k}{1 + \left(\beta_k - \gamma_k\right) \sum_{k' = 1}^K p_{k'}}   
    \end{equation}
    for ZF. Note that the ``effective SINR'' corresponds to the SNR of an additive white Gaussian noise (AWGN) channel with equivalent capacity, such that $\log_2(1 + \text{SINR})$ is an ergodic achievable rate \cite{caire2018ergodic},\cite{ngo2013energy}.     
    Further details on the derivation for \eqref{eq:SC-SINRforMRC} and \eqref{eq:SC-SINRforZF} are given in \cite{redBook}, \cite{chien2017MaMIMO}. 
    
    In a communication system each user has a quality-of-service (QoS) requirement determined by the user application and it must be satisfied by allocating system resources. QoS requirements may depend on various different parameters such as latency, rate, SINR, jitter, etc. In this work, we assume these requirements are in the form of effective SINRs, i.e., each user $k$ has a desired SINR $\alpha_k > 0$ such that the QoS requirement is 
    \begin{equation} \label{eq:SC-targetSINR}
    \text{SINR}_k \geq \alpha_k.
    \end{equation}
     This is equivalent to having a rate constraint of $\log_2(1 + \alpha_k)$. In practice, the value of $\alpha_k$ will change over time and it is important to have an efficient scheme to reallocate resources when this happens. 

    A system is called \emph{feasible} if it is possible to satisfy the SINR requirements of each user simultaneously for a given $M$ with a positive power vector $\vect{p} = [p_1, p_2, \ldots, p_k]^T$ and \emph{infeasible} otherwise. The system resources to be allocated are the number of BS antennas, $M$ and the transmit powers, $\vect{p}$. This is a generalization of prior works, in which the number of antennas is usually constant. 

Combining \eqref{eq:SC-SINRforMRC} for $k = 1, \ldots,K$ with 
\eqref{eq:SC-targetSINR}, we obtain
    \begin{equation}\label{eq:SC-MRCineq}
    	\left(\bar{M}\vect{I}_K- \vect{T}\vect{F}\right) \vect{p} \geq \bs{\nu}
    \end{equation} 
    where $\bs{\nu} = \left[\frac{\alpha_1}{\gamma_1}, \ldots, \frac{\alpha_k}{\gamma_k}\right]^T$ is the normalized noise vector, $\vect{T} = [t_{ij}] \in \mathbb{R}^{K\times K}$ is a diagonal matrix given by         
    \begin{flalign} \label{eq:SC-targetSINRmatrix} 
    t_{ij} = \begin{cases} 
    \alpha_i, & \text{if}~~ i=j, \\
     0, & \text{otherwise},
    \end{cases}
    \end{flalign}
    and $\vect{F} = [f_{ij}] \in \mathbb{R}^{K\times K}$ is a rank one matrix with
    \begin{flalign} \label{eq:SC-Chmatrix} 
    f_{ij} = \begin{cases} \frac{\beta_i}{\gamma_i}, ~~~~~~~ &\text{for MRT},  \\
    \frac{\beta_i - \gamma_i}{\gamma_i}, ~~~~~~~&\text{for ZF},
    \end{cases} 
    \end{flalign}
    and 
    \begin{flalign} \label{eq:SC-M} 
\bar{M} = \begin{cases} M, &\text{for MRT},  \\
M - K, &\text{for ZF}.
\end{cases} 
\end{flalign}
Notice that $\bar{M}$ must be a positive integer, hence for ZF precoding $M$ must be greater than the number of users. The maximum $\bar{M}$ is 
    \begin{flalign} \label{eq:MmaxSC} 
\bar{M}_{\max} = \begin{cases} M_{\max}, &\text{for MRT},  \\
M_{\max} - K, &\text{for ZF},
\end{cases} 
\end{flalign}
where $M_{\max}$ is the maximum number of antennas available at the BS.

    If \eqref{eq:SC-MRCineq} is satisfied with equality, we have 
    \begin{equation}\label{eq:SC-MRCsol} 
    \vect{p}^\dagger = \left(\bar{M}\vect{I}_K- \vect{T}\vect{F}\right)^{-1}\bs{\nu}
    \end{equation}
    which is a feasible solution (i.e., the desired SINR values can be satisfied simultaneously with a positive $\vect{p}$) if and only if the spectral radius of $\vect{T}\vect{F}$, denoted by $r(\vect{T}\vect{F})$, is less than $\bar{M}$. Furthermore, for the feasible case, the solution given by $\vect{p}^\dagger$ is the unique minimum power vector, i.e., any $\vect{p} \neq \vect{p}^\dagger$ satisfying \eqref{eq:SC-targetSINR} requires at least as much power component-wise \cite{pillai2005PF-Theorem,bambos2000chAccessALP}. 
    
    First, consider the power optimization problem 
      \begin{equation} \tag{P1}\label{pr:PR-1}
    \begin{aligned}
    & \underset{\vect{p} \succeq 0, \bar{M} \in \mathbb{Z}^+}{\text{minimize}}
    & & \sum_{k = 1}^{K} p_k \\
    & \text{subject to}
    & & \text{SINR}_k \geq \alpha_k, \\
    & & & \bar{M} \leq \bar{M}_{\max}, \\
    & & & \sum_{k = 1}^{K} p_k \leq \rho_{d},
    \end{aligned}
    \end{equation}
    where $\rho_{d}$ is the maximum transmit power of the BS. Although $\bar{M}$ is not explicitly included in the objective function, it is also an optimization parameter that affect the SINRs. 
    
    For a given $\bar{M}$, the minimum power solution to \eqref{pr:PR-1} is given by \eqref{eq:SC-MRCsol} if the system is feasible. 
    Moreover, since \eqref{eq:SC-MRCsol} provides the minimum power solution, it must be the solution to \eqref{pr:PR-1} and satisfy the power constraints. Otherwise, \eqref{pr:PR-1} has no solution with the given power constraints for the given $\bar{M}$. An important property of multi-antenna systems is that, contrary to single-antenna systems, the number of antennas is also a variable that can be optimized. We consider the problem of finding not only the optimal $\vect{p}$, but the optimal $(\bar{M},\vect{p})$-pair for \eqref{pr:PR-1}. 

    An important property of \eqref{pr:PR-1} is that even without the power constraints the feasibility of the problem is not guaranteed and depends on the number of antennas, desired SINRs, and the ratios $\frac{\beta_i}{\gamma_i} = 1 + \frac{1}{N_p\rho_{ul}\beta_k}$, which is an indicator of the channel estimation quality. The desired SINRs and $\frac{\beta_i}{\gamma_i}$ are fixed, whereas the number of BS antennas is an optimization parameter. Some results on the feasibility of the problem is provided below.    
      
    \begin{lemma}\label{lem:SC-MmaxFeasibility}
    Assume the problem defined by \eqref{pr:PR-1} is infeasible for $\bar{M} = \bar{M}_{\max}$, then there exists no $(\vect{p}, \bar{M})$ pair which provides a solution to \eqref{pr:PR-1}.  	
    \end{lemma} 
    \begin{IEEEproof}
    Rewriting \eqref{eq:SC-MRCsol}, we have 
    \begin{equation}
        \vect{p}^\dagger = \frac{1}{\bar{M}}\left(\vect{I}_K- \vect{B}(\bar{M})\right)^{-1}\bs{\nu},
    \end{equation}
    where $\vect{B}(\bar{M}) = \frac{1}{\bar{M}}\vect{T}\vect{F}$.
    	
    Let the pair $(\vect{p}', \bar{M}')$ be a feasible solution to \eqref{pr:PR-1} where $\bar{M}' < \bar{M}_{\max}$, then $r(\vect{B}(\bar{M}')) < 1 \leq r(\vect{B}(\bar{M}_{\max}))$. However, $\vect{B}(\bar{M}') \succeq \vect{B}(\bar{M}_{\max})$ which implies $r(\vect{B}(\bar{M}')) \geq  r(\vect{B}(\bar{M}_{\max}))$ \cite[Corollary 8.1.19]{hornMatrixAnalysis} and contradicts with the assumption that $(\vect{p}', \bar{M}')$ is a feasible solution.       	
    \end{IEEEproof}

    This result suggests that it is impossible to turn an infeasible system into a feasible one by reducing the number of antennas. Next, a simple extension of Lemma \ref{lem:SC-MmaxFeasibility} is stated without proof.
    \begin{lemma}\label{lem:SC-MexpandsFeasibility set}
     Let $(\vect{p}, \bar{M})$ be a pair that satisfies the SINR constraints defined in \eqref{eq:SC-MRCineq}, then there exist at least one $\vect{p}'$ vector for any $\bar{M}' > \bar{M}$ such that $(\vect{p}', \bar{M}')$ also satisfies the SINR constraints.  
    \end{lemma}
    
    The preceding analysis manifests the effect of changing the number of antennas on the feasibility of the system. However, the solution to \eqref{pr:PR-1} must minimize the transmission powers among all possible solutions in the feasible set.       
        
    \begin{theorem}\label{thm:Th-1-SC-Mmax}
    Consider the problem defined by \eqref{pr:PR-1} and assume that there exists at least one feasible solution. Let the pair $(\vect{p}^*, \bar{M}^*)$ denote the optimal solution to the problem, then $\bar{M}^* = \bar{M}_{\max}$ and $\vect{p}^* = \left(\bar{M}_{\max}\vect{I}_K- \vect{T}\vect{F}\right)^{-1}\bs{\nu}$. 
    \end{theorem}   
    \begin{IEEEproof}	
    First, we need to show that for any $(\vect{p}', \bar{M}')$ pair with $\bar{M}' < \bar{M}_{\max}$ satisfying \eqref{eq:SC-MRCineq}, we have $\vect{p}' \succeq \vect{p}^*$, where $\vect{p}^*$ is the minimum power solution corresponding to $\bar{M}^* = \bar{M}_{\max}$. Recall that for a given $\bar{M}$ the minimum power solution is given by \eqref{eq:SC-MRCsol}. Hence, it is sufficient to compare the resulting power vectors using \eqref{eq:SC-MRCsol} which satisfies the SINR constraints \eqref{eq:SC-MRCineq} with equality, i.e.,
    \begin{flalign}
    	\left(\bar{M}_{\max}\vect{I}_K- \vect{T}\vect{F}\right) \vect{p}^* = 	\left(M'\vect{I}_K- \vect{T}\vect{F}\right) \vect{p}'.
    \end{flalign}   
    Since $\bar{M}' < \bar{M}_{\max}$, the equality implies $\vect{p}' \succeq \vect{p}$. Since, with increasing $\bar{M}$, the transmission powers can be reduced and it is not possible to turn a feasible system into an infeasible one by increasing the number of antennas, as shown in Lemma \ref{lem:SC-MexpandsFeasibility set}, we have $\bar{M}^* = \bar{M}_{\max}$. Finally, the optimal power vector is given by \eqref{eq:SC-MRCsol} which concludes the proof.      	
    \end{IEEEproof}
        
        This theorem proves that one should keep all antennas on in a MaMIMO system to minimize the transmission power, which is rather intuitive given the power-scaling laws in \cite{hoydis2013howManyAntennas}.
    
    \begin{remark}
    	The solution provided by Theorem \ref{thm:Th-1-SC-Mmax} is valid for both precoding schemes, MRT and ZF. However, the exact values of the optimal solution are different since $\bar{M}$ and $\mathbf{F}$ are defined differently. Furthermore, the feasibility of one precoding scheme does not imply the feasibility of the other. 
    \end{remark}

    \begin{figure}[tb]
    	\begin{center}
    		\includegraphics[trim=2.2cm 0.1cm 0.4cm 0.1cm,clip=true,scale = 0.75]{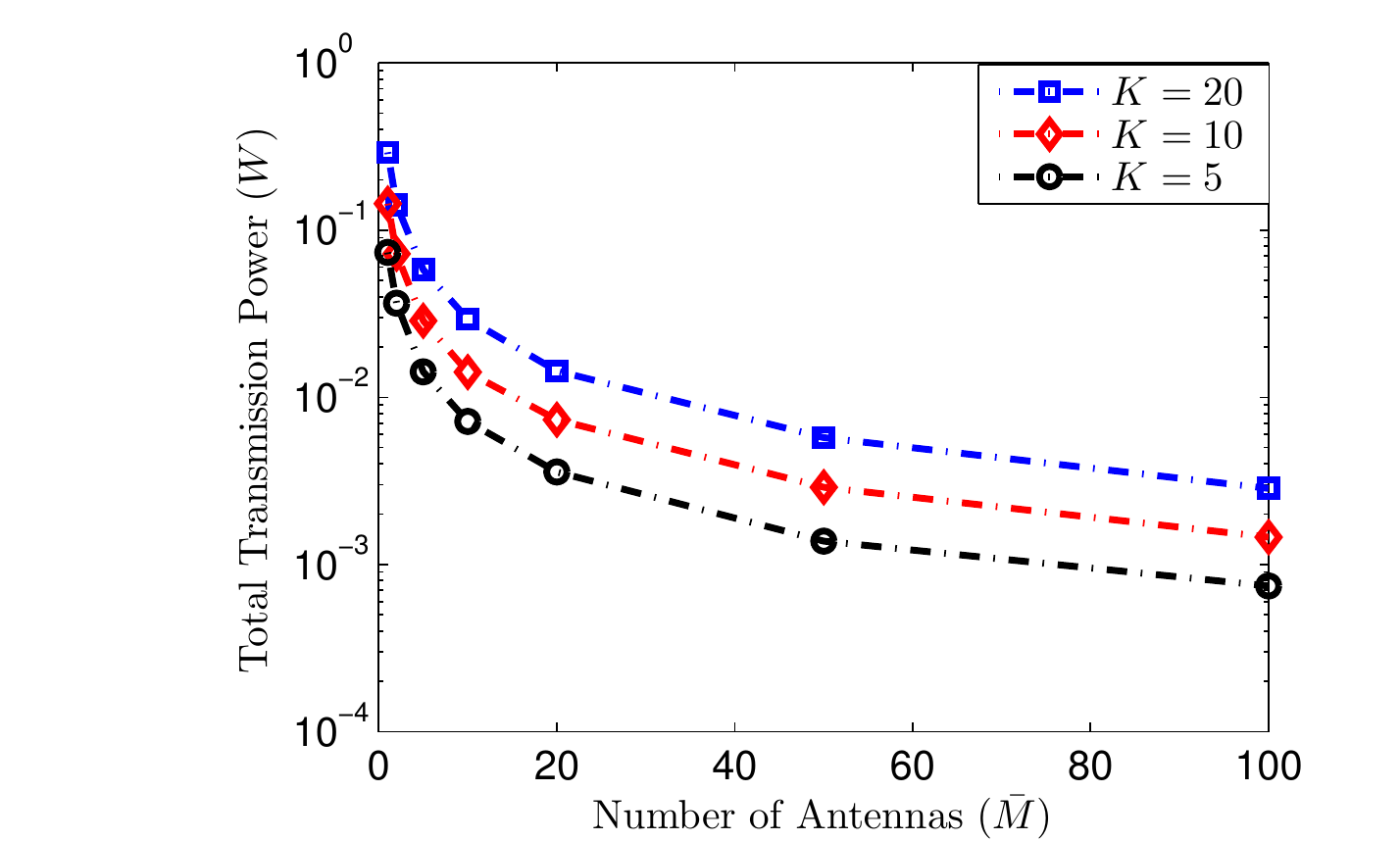}
    		\caption{Total transmission power as a function of number of BS antennas for various number of users.}
    		\label{fig:fig1}
    	\end{center}
    \end{figure}
     
         Fig.~\ref{fig:fig1} illustrates the change on the required transmission power as a function of the number of BS antennas, for different numbers of users uniformly distributed in a circular cell and ZF precoding. The simulation parameters for the numerical analysis provided throughout the paper are summarized in the Table \ref{tbl:SysParameters}. For this particular example, the SINR target is $\alpha_{k} = 1$ for each user. A low value is chosen to guarantee the feasibility of the system when the number of antennas is small. As expected, Fig.~\ref{fig:fig1} shows that the total power decreases with increasing number of antennas, which verifies the preceding analysis. The figure also shows that, as the number of users increases, the required total transmission power increases. Note that the gain from increasing $M$ is very significant when $M$ is small but diminishes as $M$ grows. Furthermore, the total transmission power grows faster than linear with respect to $K$ and decreases linearly with $M$, which is in alignment with the results provided in \cite{desset2014PowerConsumption}.

       So far, the cost of increasing the number of antennas has been neglected. In a practical system, using more antennas has a cost in terms of increased circuit power \cite{yang2013EEgreen}, \cite{ng2012ee}. Hence, a more practical consumption model is     
      \begin{equation} \label{pr:PR-2prev}
      \begin{aligned}
      c_1 M + c_2\sum_{k = 1}^{K} p_k
      \end{aligned}
      \end{equation}
      where $c_1 \geq 0$ is the circuit power consumption per antenna and $c_2 > 1$ represents the amplifier inefficiency factor which accounts for power dissipation in the amplifiers.  The values of $c_1$ and $c_2$ depends on the hardware quality deployed at the BS, operational frequency, quantization bits. The expected power consumption per antenna is $300\,$mW and the power amplifier has $30\%$ efficiency. These numbers are subject to technological scaling. Especially, the consumption per antenna reduces by a factor of two for every technology generation (every two years) \cite{desset2016eeMIMO},\cite{desset2014PowerConsumption}.

      The joint optimization problem based on the consumption model in \eqref{pr:PR-2prev} is 
      \begin{equation} \nonumber\label{pr:PR-2pre}
      \begin{aligned}
      & \underset{\vect{p} \succeq 0, \bar{M} \in \mathbb{Z}^+}{\text{minimize}}
      & & c_1 M + c_2\sum_{k = 1}^{K} p_k  \\
      & \text{subject to}
      & & \text{SINR}_k \geq \alpha_k, \\
      & & & \bar{M} \leq \bar{M}_{\max}, \\
      & & & \sum_{k = 1}^{K} p_k \leq \rho_{d}. 
      \end{aligned}
      \end{equation}    
      Consider the following equivalent optimization problem, which is obtained by dividing the cost function by $c_2$ and defining the relative power cost of operating each antenna, $c = c_1/c_2$, 
      \begin{equation} \tag{P2}\label{pr:PR-2}
      \begin{aligned}
      & \underset{\vect{p} \succeq 0, \bar{M} \in \mathbb{Z}^+}{\text{minimize}}
      & & c\bar{M} + \sum_{k = 1}^{K} p_k \\
      & \text{subject to}
      & & \text{SINR}_k \geq \alpha_k, \\
      & & & \bar{M} \leq \bar{M}_{\max}, \\
      & & & \sum_{k = 1}^{K} p_k \leq \rho_{d}. 
      \end{aligned}
      \end{equation}
      Note that we have also replaced $M$ with $\bar{M}$ in the cost function, which does not change the solution.\footnote{      Note that for ZF precoding, the first term of the cost function in \eqref{pr:PR-2} is equal to $c(M - K)$. Since $cK$ is a constant term, the problem is equivalent to the one with $cM$ and both problems have identical optimal solutions. }
      Obviously, the optimal solution depends on $c$ and \eqref{pr:PR-2} reduces to \eqref{pr:PR-1} when $c=0$.  
      
      Let 
      \begin{equation} \label{eq:SC-UtilityPr2}
      U(\vect{p},\bar{M}) = c\bar{M} + \sum_{k = 1}^{K} p_k
      \end{equation} 
      denote the cost function for a given $(\vect{p},\bar{M})$ pair.     
      Before we state our main result for single-cell systems, we first investigate the required number of antennas such that the SINR constraints can be satisfied and state the following. 
      \begin{lemma}\label{lem:SC-minM}
       Let $\bar{M}^{\text{SINR}}_{\min}$ be the minimum number of antennas such that \eqref{eq:SC-MRCsol} results in a positive power vector that satisfies the SINR constraints given in \eqref{eq:SC-targetSINR}. Then, $\bar{M}^{\text{SINR}}_{\min} = \ceil{\bar{M}^{c\text{SINR}}_{\min}}$ where
       \begin{equation} \label{eq:SC-minM}
       \bar{M}^{c\text{SINR}}_{\min} = \text{tr}(\vect{T}\vect{F}).
       \end{equation} 
      \end{lemma}
      \begin{IEEEproof}
      	Recall that \eqref{eq:SC-MRCsol} provides a feasible solution if and only if the spectral radius of $\vect{T}\vect{F}$ is less than $\bar{M}$. $\vect{T}\vect{F}$ is a rank-one matrix and $r(\vect{T}\vect{F}) = \text{tr}(\vect{T}\vect{F})$. Hence, $\bar{M}^{c\text{SINR}}_{\min} \geq r(\vect{T}\vect{F}) = \text{tr}(\vect{T}\vect{F})$.
      \end{IEEEproof}      
      Lemma \ref{lem:SC-minM} shows that the number of active antennas can be adjusted based on the traffic load of the system as $\bar{M}^{\text{SINR}}_{\min}$ depends on the large-scale coefficients, channel estimation quality and target SINR levels.

       The transmission powers are shown to decrease as the number of antennas increases in Theorem \ref{thm:Th-1-SC-Mmax}. Next, we derive the minimum number of antennas required to satisfy the both power and SINR constraints of the joint optimization problem defined in \eqref{pr:PR-2}.   
      \begin{lemma}\label{lem:SC-minMpowerSINR} Consider the problem defined in \eqref{pr:PR-2} and let $\bar{M}_{\min}$ be the minimum number of antennas required to satisfy both the SINR and power constraints. Then, $\bar{M}_{\min} = \ceil{\bar{M}^{c}_{\min}}$ where 
      \begin{equation}\label{eq:SC-minMconstraints}
      \bar{M}^c_{\min} = \frac{1}{\rho_{d}}\sum_{k' = 1}^K \frac{\alpha_{k'}}{\gamma_{k'}} + \text{tr}(\vect{T}\vect{F}). 
       \end{equation} 
      \end{lemma}     
      \begin{IEEEproof}
           The power constraint can be written as
           \begin{eqnarray}
           \rho_{d} &\geq& \sum_{k = 1}^{K} p_k \\
           &=& \vect{1}^T\vect{p} \\ 
           &=& \vect{1}^T \left(\bar{M}\vect{I}_K- \vect{T}\vect{F}\right)^{-1}\bs{\nu} \label{eq:pProof-2} \\ &=& \frac{1}{\bar{M}}\vect{1}^T \left(\vect{I}_K +  \frac{1}{\bar{M} - \text{tr}(\vect{T}\vect{F})}\vect{T}\vect{F}\right)\bs{\nu} \label{eq:pProof-3}
           \end{eqnarray}
            where \eqref{eq:pProof-2} and \eqref{eq:pProof-3} follows from \eqref{eq:SC-MRCsol} and Lemma \ref{lem:SMV}, respectively. Finally \eqref{eq:SC-minMconstraints} can be obtained by solving for $\bar{M}$ in \eqref{eq:pProof-3}.                 
            Note that $\bar{M}^c_{\min} >  \text{tr}(\vect{T}\vect{F})$ as $\alpha, \gamma$ and $\rho_{d}$ are all strictly positive variables. Hence, we have $\bar{M}^c_{\min} > \bar{M}^{c\text{SINR}}_{\min}$ which implies $\bar{M}_{\min} \geq \bar{M}^{\text{SINR}}_{\min}$ and $\bar{M}_{\min} = \ceil{\bar{M}^{c}_{\min}}$.               
           \end{IEEEproof}
      
      Lemma \ref{lem:SC-minMpowerSINR} reveals the minimum required number of antennas such that \eqref{pr:PR-2} has a solution. This provides a lower bound on the optimum number of antennas.  
      The main result for the single-cell systems regarding \eqref{pr:PR-2} is stated below. 
      
           \begin{theorem} \label{thm:thm-2-SC-optM}
      	Assume that there exists at least one $(\vect{p}, \bar{M}')$-pair that satisfies the constraints in \eqref{pr:PR-2} with $\bar{M}' \leq \bar{M}_{\max}$ and 
      	let $(\vect{p}^*, \bar{M}^*)$ denote the optimal solution with $\bar{M}^*\in \mathbb{R}^+$, then 
      	\begin{equation}
      	\bar{M}^* = \min(\bar{M}_{\max},\max(\bar{M}^\dagger,\bar{M}_{\min})),
      	\end{equation}
      	 where 
      	      	\begin{flalign} \label{eq:SC-optM-MRC}
      	      	\bar{M}^\dagger =  \text{tr}(\vect{T}\vect{F}) + \left(\frac{1}{c}\sum_{k' = 1}^K \frac{\alpha_{k'}}{\gamma_{k'}}\right)^{1/2}
      	      	\end{flalign}     
      	and 
      	\begin{equation}\label{eq:SC-optPower}
      	\vect{p}^* = \frac{1}{\bar{M}^*} \left(\vect{I}_K +  \frac{1}{\bar{M}^* - \text{tr}(\vect{T}\vect{F})}\vect{T}\vect{F}\right)\bs{\nu}.   
      	\end{equation}
      	
      \end{theorem}
      \begin{IEEEproof}
      	First, note that for a given $\bar{M} \geq \bar{M}_{\min}$, the transmit powers that minimize $U(\vect{p},\bar{M})$ are given by \eqref{eq:SC-MRCsol}. Hence, $\tilde{U}(\bar{M}) = U(\vect{p},\bar{M})$ can be considered a function of only $\bar{M}$, given by
      	\begin{eqnarray} 
      	\tilde{U}(\bar{M}) &=& c\bar{M} + \vect{1}^T \vect{p}\nonumber \\ 
      	&=& c\bar{M} + \vect{1}^T \left(\bar{M}\vect{I}_K- \vect{T}\vect{F}\right)^{-1}\bs{\nu}. \label{eq:SC-convexity-1}
      	\end{eqnarray}
      	Using \eqref{eq:SMV-BM}, \eqref{eq:SC-convexity-1} can further be simplified as
      	\begin{eqnarray}
      	\tilde{U}(\bar{M}) &=& c\bar{M} + \frac{1}{\bar{M}}\vect{1}^T \left(\vect{I}_K +  \frac{1}{\bar{M} - \text{tr}(\vect{T}\vect{F})}\vect{T}\vect{F}\right)\bs{\nu} \nonumber \\  
      	&=& c\bar{M} + \frac{1}{\bar{M}} \left( \tau+ \frac{\text{tr}(\vect{T}\vect{F})}{\bar{M} - \text{tr}(\vect{T}\vect{F})} \tau\right) \label{eq:SC-convexity-2}
      	\end{eqnarray}
      	where 
      	\begin{equation}\label{eq:SC-tau}
      	\tau = \vect{1}^T \bs{\nu} =  \sum_{k' = 1}^K \frac{\alpha_{k'}}{\gamma_{k'}}.
      	\end{equation}
      	It is straightforward to show that $\tilde{U}(\bar{M})$ is a strictly convex function with respect to $\bar{M}$ by examining the second derivative of \eqref{eq:SC-convexity-2}. Therefore, the optimal solution can be found by taking the derivative of $\tilde{U}(\bar{M})$ with respect to $\bar{M}$ and equating it to zero. Let $\mu = \text{tr}(\vect{T}\vect{F})$, then
      	\begin{equation}\label{eq:SC-derObjectiveFncWrtM}
      	\frac{d\,\tilde{U}(\bar{M})}{d\,\bar{M}} = c - \frac{\tau}{\bar{M}^2} - \frac{\mu\tau (2\bar{M} - \mu)}{\bar{M}^2(\bar{M}-\mu)^2}
      	\end{equation} 
      	which should be equal to zero when $\bar{M} = M^\dagger$. Rearranging the terms we get
      	\begin{eqnarray}
      	c\bar{M}^4 - 2\mu c\bar{M}^3 + (c\mu^2 - \tau)\bar{M}^2 &=& 0  \label{eq:SC-P2der}\\
      	\left( \bar{M} - \mu\right)^2 &=& \frac{\tau}{c}. \label{eq:SC-P2der2}
      	\end{eqnarray}
      	In \eqref{eq:SC-P2der}, two zero roots are discarded as they correspond to $\bar{M} = 0$. From \eqref{eq:SC-P2der2}, we get the possible solutions $\bar{M} = \mu \pm \sqrt{\frac{\tau}{c}} $ and the optimal solution is 
      	\begin{flalign} \label{eq:SC-optM-P2}
      	\bar{M}^\dagger =\mu + \sqrt{\frac{\tau}{c}}
      	\end{flalign}  
      	as $\bar{M} = \mu - \sqrt{\frac{\tau}{c}}$ corresponds to an infeasible system.
      	
      	The optimal solution given by $\bar{M}^\dagger$ must satisfy the constraints and therefore must be greater than $\bar{M}_{\min}$.  For the case where $\max(\bar{M}_{\min},\bar{M}^\dagger) > \bar{M}_{\max}$, the optimal solution is clearly $\bar{M}^* = \bar{M}_{\max}$ as $\tilde{U}(\bar{M})$ is a convex function of $\bar{M}$.    
      \end{IEEEproof}
         \begin{remark}
  	In Theorem \ref{thm:thm-2-SC-optM}, the integer constraint on $\bar{M}$ is relaxed and the resulting optimal $\bar{M}$ is not necessarily an integer. However, it is clear that the optimal $\bar{M}$ is either  $\floor{\bar{M}}$ or  $\ceil{\bar{M}}$ as a result of the convexity of the cost function. 
  \end{remark}
  
       \begin{figure}[tb]
   	    \begin{center}
   		\includegraphics[trim=0cm 0.1cm 0.4cm 0.1cm,clip=true,width = 8cm]{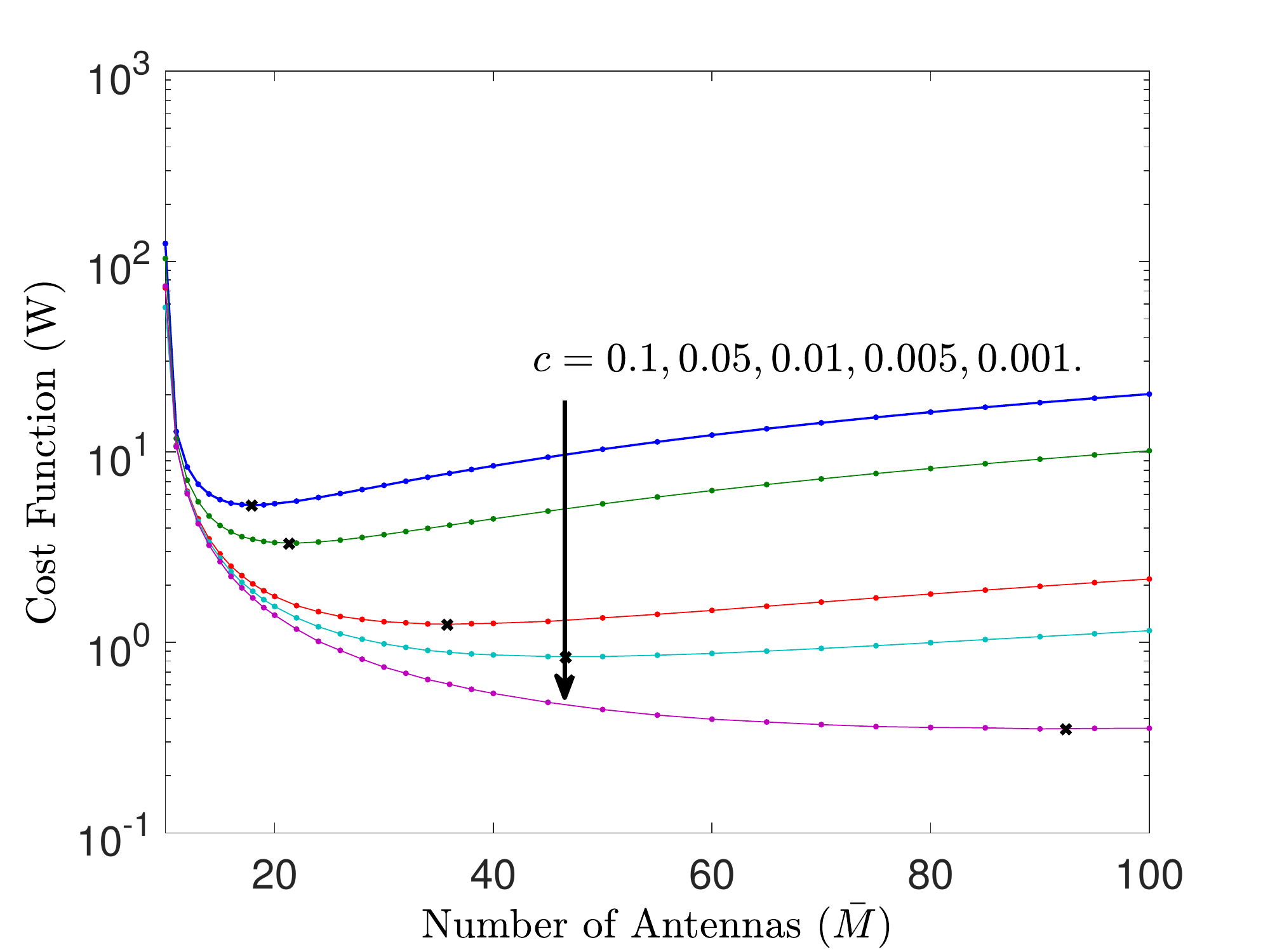}
   		\caption{Cost function as a function of number of BS antennas for $10$ users with various $c$ values and MRT precoding. For each curve, the black $\boldsymbol{\times}$ represent the optimal cost function value obtained by $M^\dagger$ defined by \eqref{eq:SC-optM-P2}.}
   		\label{fig:fig2}
   	     \end{center}
      \end{figure}
                     
      In Fig.~\ref{fig:fig2}, power consumption for various $c$ values and MRT precoding ($\bar{M} = M$) are depicted. In each trial, the optimal power vector for the given $M$ is obtained using \eqref{eq:SC-MRCsol} and $U(\vect{p},M)$ is computed. Each point represents the average of $5000$ independent trials with uniformly random user locations and the crosses ($\boldsymbol{\times}$) represent the average of $M^\dagger$ computed by \eqref{eq:SC-optM-P2} and the corresponding $U(\vect{p},M^\dagger)$. For each $c$, $U(\vect{p},M^\dagger)$ is less than the minimum obtained. However, an important point is that $M^\dagger$ is not necessarily an integer whereas the curves are obtained by using integer $M$ values. As $c$ decreases, the resulting cost functions decrease and the minimum is attained at a higher $M$ as expected.       
    It is only when $c$ is small that it is optimal to turn all antennas on, while in other cases one can save energy by turning antennas off.
        
    Next, we investigate the relation between 
    the optimum $\bar{M}$ for MRT and ZF.
    
    \begin{lemma} \label{lem:SC-optMdiffMRC-ZF}
    	Consider the problem defined by \eqref{pr:PR-2} and let $\bar{M}^\dagger_\mathrm{MRT}$ and $\bar{M}^\dagger_\mathrm{ZF}$ denote the optimum $\bar{M} \in \mathbb{R}^+$ for MRT and ZF processing such that $\bar{M}_{\min} \leq \bar{M} \leq \bar{M}_{\max}$, respectively. Then,
    	\begin{equation} \label{eq:SC-diffoptMdiffMRC-ZF}
    	\bar{M}^\dagger_\mathrm{MRT} - \bar{M}^\dagger_\mathrm{ZF} = \sum_{k' = 1}^K\alpha_{k'} - K.    
    	\end{equation}    	       
    \end{lemma}
    \begin{IEEEproof}
    	The difference $\bar{M}^\dagger_\mathrm{MRT} - \bar{M}^\dagger_\mathrm{ZF}$ can be computed by substituting the definitions of $\vect{F}$ and $\bar{M}$ for MRT and ZF from \eqref{eq:SC-Chmatrix} and \eqref{eq:SC-M} into \eqref{eq:SC-optM-MRC}, which gives
    	\begin{eqnarray} \nonumber
    	\bar{M}^\dagger_\mathrm{MRT} - \bar{M}^\dagger_\mathrm{ZF} =  \sum_{k' = 1}^K\alpha_{k'} \frac{\beta_{k'}}{\gamma_{k'}} - 
    	\left(\sum_{k' = 1}^K\alpha_{k'}\left( \frac{\beta_{k'}}{\gamma_{k'}} -1\right)\right) - K.      
    	\end{eqnarray}
    	This can be simplified to obtain \eqref{eq:SC-diffoptMdiffMRC-ZF}. 
    \end{IEEEproof} 
    
    For the case when all users have the same SINR requirements, i.e., $\alpha_k = \alpha$ for all $k$, \eqref{eq:SC-diffoptMdiffMRC-ZF} reduces to $\bar{M}^\dagger_\mathrm{MRT} - \bar{M}^\dagger_\mathrm{ZF} = K (\alpha - 1)$. If $\alpha = 1$, then the optimal number of antennas is equal for both approaches assuming that the optimal number of antennas is smaller or equal to $\bar{M}_{\max}$. MRT requires more antennas for $\alpha > 1$ and vice versa.
    
\begin{figure}[t]
    	\begin{center}
    		\includegraphics[trim=2cm 0.1cm 0.4cm 0.1cm,clip=true,scale = 0.6]{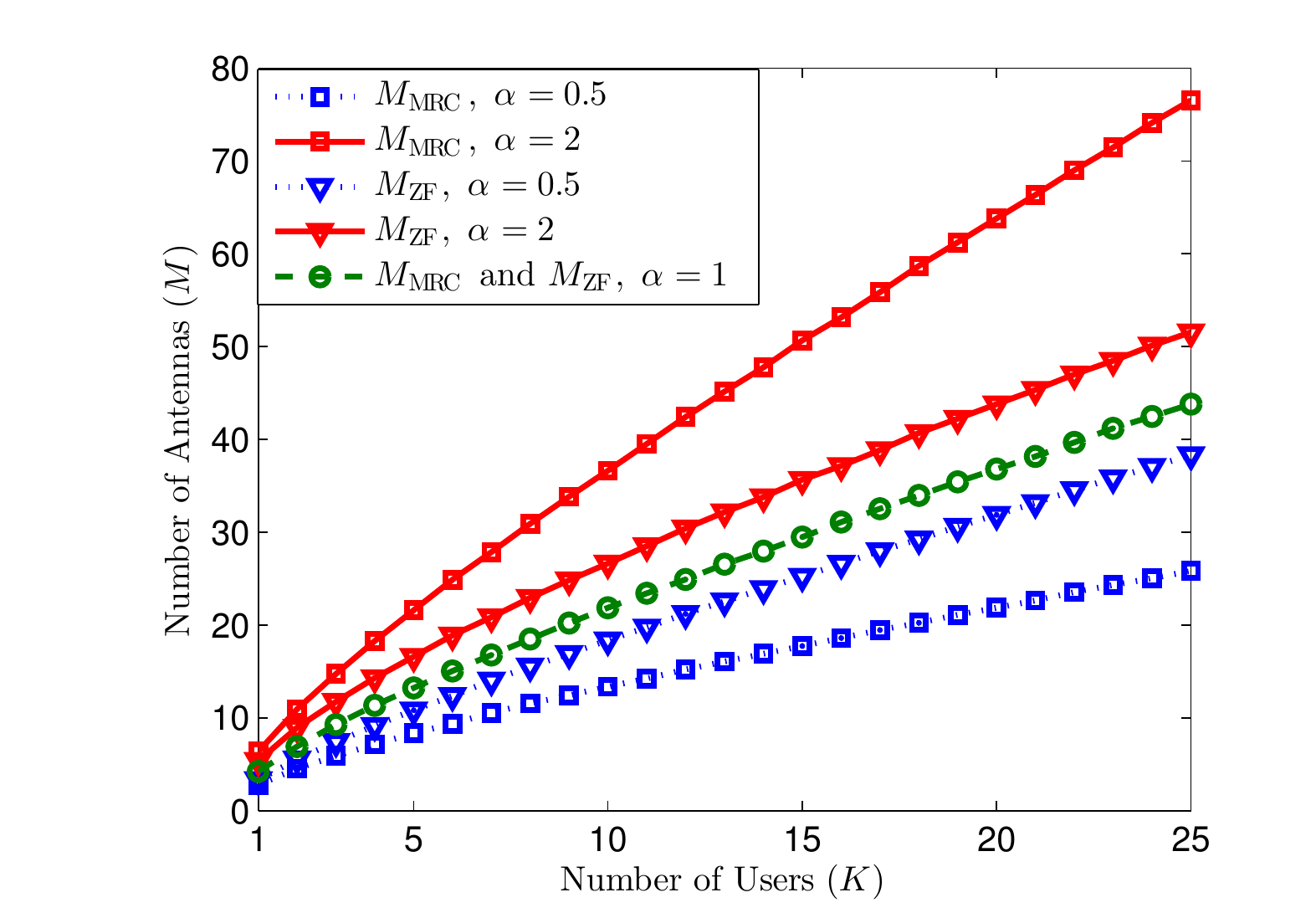}
    		\caption{Comparison of the optimal number of antennas for \eqref{pr:PR-2} with MRT and ZF as a function of number of users. }
    		\label{fig:fig3}
    	\end{center}
    	\vspace{-0.5cm}
    \end{figure}

    Fig.~\ref{fig:fig3} illustrates the optimum number of antennas for \eqref{pr:PR-2} for MRT and ZF processing with $c = 0.001$. The integer constraint is relaxed to obtain the smooth curves depicted in the figure. The optimal number of antennas for MRT is lower when $\alpha < 1$, it is lower for ZF when $\alpha > 1$, and equal when $\alpha = 1$, which is in alignment with Lemma \ref{lem:SC-optMdiffMRC-ZF}. 
    In all of the considered cases, it is optimal to turn off a subset of the antennas. If MaMIMO is defined as a system with at least 50 active antennas, then we need to have tens of users and/or high QoS constraints to make MaMIMO optimal.
    
    Note that \eqref{eq:SC-diffoptMdiffMRC-ZF} allows us to also compare the two precoding techniques in a setup with heterogeneous SINR requirements and find the optimal number of antennas. It can be concluded that if the average SINR target is smaller than 1, MRT results in fewer required antennas at the optimal solution and vice versa. 
    
    \section{Multi-Cell System} \label{sec:MC system}
    In this section, we extend the single-cell analysis in the previous section to a multi-cell setting, where both inter-cell and intra-cell interference are considered. Furthermore, it is usually not feasible to assign orthogonal pilot sequences to every user in the system, which results in \emph{pilot contamination} that deteriorates the channel estimation quality and leads to coherent interference. We consider a setup where each BS serves  $K$ users and each user is associated with a single BS. We assume that the users in a cell have mutually orthogonal pilots while the pilot sequences of two distinct cells are either completely orthogonal or replicated. The cells which assign identical pilot sequences to their users causes pilot contamination to each other. The pilot contaminating cells cause coherent interference that grows with the number of antennas, which is a crucial difference from single-cell systems where the interference terms are independent of the number of antennas.
    
    \subsection{Uplink Training and Downlink Data Transmission}
    We adopt the notation from \cite{redBook}. Let $\vect{g}^j_{lk}$ denote the channel between user $k$ in cell $l$ and BS $j$.
     Then, BS $l$ receives the signal
    \begin{equation}\label{eq:MC-receivedSig1}
        \vect{Y}_{l} =  \sqrt{N_p \rho_{ul}} \sum_{l' = 1}^L\sum_{k' = 1}^{K}\vect{g}_{l'k'}^{l}\bs{\varphi}_{l'k'}^H + \vect{Z}
    \end{equation}
    where $\bs{\varphi}_{lk}$ denotes the pilot sequence of user $k$ in cell $l$. In order to estimate $\vect{g}^l_{lk}$, BS $l$ performs a de-spreading process on the received signal as follows
    \begin{eqnarray} 
                 \vect{y}'_{lk}  &=& \vect{Y}_{l}\bs{\varphi}_{lk} \nonumber \\ 
     &=&  \sqrt{N_p \rho_{ul}} \sum_{l' = 1}^L\sum_{k' = 1}^{K}\vect{g}_{l'k'}^{l}\bs{\varphi}_{l'k'}^H\bs{\varphi}_{lk} + \vect{z}'. \label{eq:MC-depreadingRecSig}
    \end{eqnarray}   
	The MMSE estimate of $\vect{g}^l_{lk}$ based on \eqref{eq:MC-depreadingRecSig}, is denoted 
	by $\hat{\vect{g}}^l_{lk}$ and contains $M_l$ identically distributed components $\mathsf{CN}(0,\gamma^l_{lk})$ with
		\begin{eqnarray}
	\gamma^l_{lk} 
	&=& \frac{N_p \rho_{ul}\left(\beta^l_{lk}\right)^2 }{1 + N_p \rho_{ul} \sum_{l' = 1}^L\sum_{k' = 1}^{K}\beta_{l'k'}^{l}|\bs{\varphi}_{l'k'}^H\bs{\varphi}^{}_{lk}|^2},
	\end{eqnarray}
	where  $\beta^j_{lk}$ denote the large-scale fading coefficient \cite{Kay1993estimation}.
		Note that the summation term in the denominator contains only the terms originating from the users with identical pilot sequences. 
	The channel estimation error with MMSE,  $\tilde{\vect{g}}^l_{lk} = \vect{\hat{g}}^l_{lk} - \vect{g}^l_{lk}$, is independent of the estimate and its elements are i.i.d. with $\mathsf{CN}(0,\beta^l_{lk} - \gamma^l_{lk})$.

	Similar to the single-cell case, we consider MRT and ZF precoding. 	Let $\mathcal{P}_l$ denote the set of cells that share the same pilot sequences with cell $l$ including the own cell, then
	  the effective SINR for MRT is \cite{redBook}, \cite{ashikhmin2012pilot}, \cite{yang2013total}, \cite{chien2017MaMIMO},  
	\begin{equation}\label{eq:MC-SINR-MRC}
\mathrm{SINR}_{lk}^{\mathrm{MRT}} = \frac{M_l\gamma^l_{lk}p_{lk}}{1 + \sum\limits_{l' = 1}^L\sum\limits_{k' = 1}^Kp_{l'k'}\beta^{l'}_{lk} + \sum\limits_{l' \in \mathcal{P}_l\backslash \{l\}}M^{}_{l'}\gamma^{l'}_{lk}p^{}_{l'k}}
	\end{equation}
	 where  $p_{lk}$ is the normalized power allocation of BS $l$ for user $k$ and we assume that in cells with shared pilots $k$th users have identical pilot sequences. Similarly for ZF, the effective SINR is given by \eqref{eq:MC-SINR-ZF}.

	 \begin{figure*}[!t]
	 	\normalsize
	 	
	 	\begin{equation}
	 	\label{eq:MC-SINR-ZF}
	 	\mathrm{SINR}_{lk}^{\mathrm{ZF}} = 
	 	\frac{\left(M_l-K\right)\gamma^l_{lk}p_{lk}}{1 + \sum\limits_{k' = 1}^K\left( \sum\limits_{l' \in \mathcal{P}_l}p_{l'k'}\left(\beta^{l'}_{lk}- \gamma^{l'}_{lk}\right) + \sum\limits_{l' \notin P_l}p_{l'k'}\beta^{l'}_{lk}\right)+  \sum\limits_{l' \in P_l\backslash \{l\}}(M_{l'}-K)\gamma^{l'}_{lk}p_{l'k}}
	 	\end{equation}
	 	\hrulefill
	 \end{figure*}
	 	 
	 First, we consider a multi-cell system with MRT and start with the following minimization problem: 
	   
	 \begin{equation} \tag{P3}\label{pr:PR-3}
	 \begin{aligned}
	 & \underset{\vect{p}_1,\ldots,\vect{p}_L\succeq 0, \vect{m} \in \mathbb{Z}^+}{\text{minimize}}
	 & & \sum_{l = 1}^L\sum_{k = 1}^{K} p_{lk} \\	 & ~~~~\text{subject to}
	 & & \text{SINR}_{lk} \geq \alpha_{lk},~l = 1, \ldots, L,~k = 1,\ldots, K\\
	 & & & M_l \leq M_{max},~l = 1, \ldots, L \\
	 & & & \sum_{k = 1}^{K} p_{lk} \leq \rho_{d},~l = 1, \ldots, L.
	 \end{aligned}
	 \end{equation}
	 where $\vect{p}_l = [p_{l1}, \ldots, p_{lK}]^T$ denotes the transmission power vector for BS $l$ and $\vect{m} = [M_1,\ldots, M_L]^T$ is the vector with the number of active antennas at each BS. Note that we have assumed the same power constraint for all BSs and the power consumption of each BS contribute equally to the objective function. The QoS requirement of user $k$ in cell $l$ is $\alpha_{lk} > 0$.
	 
	 Combining the SINR constraints 
	 \begin{equation} \label{eq:MC-SINRconstraints}
	 \text{SINR}_{lk} \geq \alpha_{lk}
	 \end{equation}
	 with \eqref{eq:MC-SINR-MRC} and \eqref{eq:MC-SINR-ZF},
	 and considering all cells in the system, we obtain 
	 \begin{equation}
	 \left(\vect{M} - \vect{T}\vect{F}-\vect{T}\vect{A}\vect{M}\right)\vect{p}\geq \boldsymbol{v}
	 \end{equation}
	 in vector notation. Here $\vect{p} = [\vect{p}_1, \ldots, \vect{p}_L]^T$ is the power vector, $\boldsymbol{v} = \left[\frac{\alpha_{11}}{\gamma^1_{11}}, \frac{\alpha_{12}}{\gamma^1_{12}}, \ldots, \frac{\alpha_{LK}}{\gamma^L_{LK} }\right]^T$ is the normalized noise vector,
	 \begin{equation} \label{eq:MC-antennaMatrix}
\vect{M} = 	\left[\begin{array}{cccc}
	\vect{M}_1& & &\\
	 &\vect{M}_2&\multicolumn{2}{c}{\smash{\raisebox{.5\normalbaselineskip}{$\vect{0}$}}}\\
	 & &\ddots &\\
	\multicolumn{2}{c}{\smash{\raisebox{.5\normalbaselineskip}{$\vect{0}$}}}& &\vect{M}_L\\
	\end{array}\right]
	 \end{equation}
	 with each $\vect{M}_i = \bar{M}_i\vect{I}_K$ is a diagonal matrix where
	  \begin{flalign} 
	 \bar{M}_i = \begin{cases} 
	 M_i, & \text{for MRT}, \\
	 M_i - K, & \text{for ZF},
	 \end{cases}
	 \end{flalign}
	 $\vect{T} = [t_{ij}] \in \mathbb{R}^{LK\times LK}$ is a diagonal matrix with 
	 \begin{flalign} 
	 t_{ij} = \begin{cases} 
	 \alpha_{i'j'}, & \text{if}~~ i=j, \\
	 0, & \text{otherwise},
	 \end{cases}
	 \end{flalign}
	 where $i' = \ceil{\frac{i}{K}}$ and $ j' = (j-1)\,\text{mod}\,K + 1$; $\vect{F}~=~[f_{ij}]~\in~\mathbb{R}^{LK\times LK}$ with
	     \begin{flalign} 
	 f_{ij} =  
	 \frac{\beta^{j'}_{i'k'}}{\gamma_{i'k'}^{i'}},~~~~\forall i,~j,
	 \end{flalign}
	  for MRT and 
	  \begin{flalign} 
	  f_{ij} =\begin{cases}  
	  \frac{\beta^{j'}_{i'k'}}{\gamma_{i'k'}^{i'}},&~~~\text{if}~ j' \notin \mathcal{P}_{i'},\\
	  \frac{\beta^{j'}_{i'k'} - \gamma^{j'}_{i'k'}}{\gamma_{i'k'}^{i'}},&~~~\text{if}~ j' \in \mathcal{P}_{i'},
	  \end{cases}
	  \end{flalign}
	  for ZF. Here, $i' = \ceil{\frac{i}{K}}$, $j' = \ceil{\frac{j}{K}}$ and $ k' = (i-1)\,\text{mod}\,K + 1$; $\vect{A} = [a_{ij}] \in \mathbb{R}^{LK\times LK}$ with
	 \begin{flalign} 
	 a_{ij} = \begin{cases} 
	 \frac{\gamma_{i'k''}^{j'}}{\gamma_{i'k'}^{i'}}, & \text{if}~~ k' = k''~~\text{and}~~j' \in \mathcal{P}_{i'} \backslash\{i'\}, \\
	 0, & \text{otherwise},
	 \end{cases}
	 \end{flalign}
	where $i' = \ceil{\frac{i}{K}}$, $j' = \ceil{\frac{j}{K}}$, $k' = (i -1)\,\text{mod}\,K + 1$ and $k'' = (j -1)\,\text{mod}\,K + 1$ for both MRT and ZF.
		   
	It is straightforward to show that the optimal solution to \eqref{pr:PR-3} is obtained when the SINR constraints are satisfied with equality. Hence, the minimum power solution is
	\begin{equation}\label{eq:MC-paretoOptimalSolution}
	 \vect{p} =  \left(\vect{M} - \vect{T}\vect{F}-\vect{T}\vect{A}\vect{M}\right)^{-1}\bs{v},
	\end{equation}  
	if and only if the spectral radius of $ \vect{T}\vect{F}\vect{M}^{-1}+\vect{T}\vect{A}$ is less than unity. The solution in \eqref{eq:MC-paretoOptimalSolution} guarantees that the SINR constraints are satisfied for all users. Furthermore, it is the minimum power solution, i.e., any other solution satisfying \eqref{eq:MC-SINRconstraints} requires at least as much power component-wise \cite{bambos2000chAccessALP}. This implies that if the solution given in \eqref{eq:MC-paretoOptimalSolution} does not satisfy the power constraints in \eqref{pr:PR-3}, there exist no other power vector that satisfies the SINR and power constraints in \eqref{pr:PR-3}, simultaneously. Contrary to the single-cell case, in the multi-cell case there is a coherent interference term in \eqref{eq:MC-SINRconstraints} that scales with the number of BS antennas in the pilot-sharing cells which limits the SINRs of the system. Before we investigate the effect of coherent interference on the system and the interplay between the number of BS antennas and transmission powers, a brief introduction to M-matrices is provided. 
	\begin{defn}
		Let $\vect{A}$ be a square matrix with nonpositive off-diagonal elements and nonnegative diagonal entries, then $\vect{A}$ can be expressed as 
		\begin{equation} \label{eq:M-matrix}
		\vect{A} = s\vect{I} - \vect{B},
		\end{equation}
		where $s > 0$ and $\vect{B}$ is a nonnegative matrix. Furthermore, assume that $s > r(\vect{B})$, then $\vect{A}$ is a nonsingular M-matrix \cite{berman1994NonnegativeMatrixBook}.   
	\end{defn}
	Next, we summarize some of the properties of M-matrices which will be utilized in the succeeding analysis.
	\begin{lemma}Let $\vect{A}$ be a square matrix, then the following statements are equivalent \cite[Chapter 6]{berman1994NonnegativeMatrixBook}:
	\begin{itemize}
		\item $\vect{A}$ is a nonsingular M-matrix.
		\item $\vect{A}^{-1}$ exists and $\vect{A}^{-1} \succeq \vect{0}$.
		\item $\vect{A}$ is a monotone matrix, i.e., for all real vectors $\vect{x}$, $\vect{A}\vect{x} \succeq \vect{0} \Rightarrow \vect{x} \succeq \vect{0}$.		 
  	\end{itemize}
	\end{lemma} 
    The feasibility conditions based on the spectral radius of $\vect{T}\vect{F}$ in single-cell case and $\vect{T}\vect{F} + \vect{T}\vect{A}\vect{M}$ in multi-cell case actually correspond to the resulting matrix being an M-matrix. 

    In order to obtain the solution to \eqref{pr:PR-3}, we need to investigate the interplay between the number of antennas and the transmission powers. Although it is clear that increasing the number of antennas in cell $l$ will result in a smaller transmission power vector $\vect{p}_l$ for cell $l$, the effect on the overall system is not clear, since the number of antennas also appear in the coherent interference term. Next, we address this problem and state the following.      
	\begin{lemma} \label{lem:MC-diffMdiffP}
	  Consider the problem defined by \eqref{pr:PR-3} and let $(\vect{m}_1,\vect{p}_1)$, $(\vect{m}_2,\vect{p}_2)$ be two solutions satisfying 
	  \eqref{eq:MC-SINRconstraints} with equality. If $\vect{m}_1 \succeq \vect{m}_2$, then $\vect{p}_1 \preceq \vect{p}_2$.
	\end{lemma}

    \begin{IEEEproof}
 	  Since the SINR constraints are satisfied with equality, we have
 \begin{equation}\label{eq:MC-diffMequivalance}
 \left(\vect{M}_{\vect{m}_1}- \vect{T}\vect{F}-\vect{T}\vect{A}\vect{M}_{\vect{m}_1}\right)\vect{p}_1 = \left(\vect{M}_{\vect{m}_2} - \vect{T}\vect{F}-\vect{T}\vect{A}\vect{M}_{\vect{m}_2}\right)\vect{p}_2	
 \end{equation}
 where $\vect{M}_{\vect{m}_1}$, $\vect{M}_{\vect{m}_2}$ are the antenna matrices defined in \eqref{eq:MC-antennaMatrix} for the corresponding $\vect{m}_1$ and $\vect{m}_2$. Let $\vect{N} = \vect{M}_{\vect{m}_1} - \vect{M}_{\vect{m}_2}$ be a diagonal matrix denoting the difference. It is a non-negative matrix since $\vect{m}_1 \succeq \vect{m}_2$ and \eqref{eq:MC-diffMequivalance} can be simplified as  
 \begin{flalign}\label{eq:MC-diffMequivalance-2}
\left(\vect{I}_{LK} - \vect{T}\vect{F}\vect{M}_{\vect{m}_2}^{-1} - \vect{T}\vect{A}\right)^{-1}\left(\vect{I}_{LK} - \vect{T}\vect{A}\right)\vect{N}\vect{p}_1 = \nonumber \\  \vect{M}_{\vect{m}_2}\left(\vect{p}_2 - \vect{p}_1\right).
 \end{flalign}  
 The left-hand side of \eqref{eq:MC-diffMequivalance-2} is non-negative since 
 \begin{flalign}
  \left(\vect{I}_{LK} - \vect{T}\vect{F}\vect{M}_{\vect{m}_2}^{-1} - \vect{T}\vect{A}\right)^{-1}\left(\vect{I}_{LK} - \vect{T}\vect{A}\right) =\nonumber \\ \vect{I}_{LK} + \left(\vect{I}_{LK} - \vect{T}\vect{F}\vect{M}_{\vect{m}_2}^{-1} - \vect{T}\vect{A}\right)^{-1}\vect{T}\vect{F}\vect{M}_{\vect{m}_2}^{-1} 
 \end{flalign}
 and $\left(\vect{I}_{LK} -  \vect{T}\vect{F}\vect{M}_{\vect{m}_2}^{-1} -\vect{T}\vect{A}\right)$ is an M-matrix. 
  Then, the nonnegativity of left-hand side of \eqref{eq:MC-diffMequivalance-2} implies that $\vect{p}_2 - \vect{p}_1 \succeq \vect{0}$. 
 	\end{IEEEproof}
     Lemma \ref{lem:MC-diffMdiffP} reveals that the total transmission power can be reduced by increasing the number of antennas. Furthermore, increasing the number of antennas in one cell also results in a lower total transmission power. This shows that even in the presence of coherent interference the overall system performance can be improved by increasing the number of antennas of BSs.   
     \begin{lemma}\label{lem:MC-infeas}
        Let $(\vect{m}_1,\vect{p}_1)$ be a pair satisfying 
     	\eqref{eq:MC-SINRconstraints}. Then, there exists at least one $\vect{p}'$ for any $\vect{m}' \succeq \vect{m}_1$ such that $(\vect{m}',\vect{p}')$ also satisfies \eqref{eq:MC-SINRconstraints}.
     \end{lemma}
 \begin{IEEEproof}
 	Note that $\vect{T}, \vect{F}, \vect{A}$ are all non-negative matrices and 
 	    \begin{equation}
 	 \vect{T}\vect{F}\vect{M}_{\vect{m}'}^{-1} + \vect{T}\vect{A} \preceq \vect{T}\vect{F}\vect{M}_{\vect{m}_1}^{-1} + \vect{T}\vect{A}.
 	 \end{equation} 
 Since $(\vect{m}_1,\vect{p}_1)$ satisfies \eqref{eq:MC-SINRconstraints}, using \cite[Corollary 8.1.19]{hornMatrixAnalysis}, we can obtain
   \begin{equation}
 r(\vect{T}\vect{F}\vect{M}_{\vect{m}'}^{-1} + \vect{T}\vect{A}) \leq r(\vect{T}\vect{F}\vect{M}_{\vect{m}_1}^{-1} + \vect{T}\vect{A}) < 1.
 \end{equation}
 which concludes the proof. 
 \end{IEEEproof}

 Lemma \ref{lem:MC-infeas} shows that it is not possible to turn a feasible system into an infeasible one by increasing the number of antennas, similar to single-cell case. Furthermore, combined with Lemma \ref{lem:MC-diffMdiffP}, a system with increased number of antennas will also require less transmission power. Unfortunately, Lemmas \ref{lem:MC-diffMdiffP} and \ref{lem:MC-infeas} do not imply that a feasible solution can always be found by increasing the number of BS antennas. Next, we consider the feasibility of the system with respect to the number of antennas and state the following.   
 
	\begin{lemma}\label{lem:MC-feasibility}
		Consider the problem defined by \eqref{pr:PR-3} and assume that $r(\vect{T}\vect{A}) \geq 1$, then, there exist no $(\vect{m},\vect{p})$ pair that satisfies the SINR constraints given in \eqref{eq:MC-SINRconstraints}. 
	\end{lemma}
	  \begin{IEEEproof}
      Recall that the SINR constraints given in \eqref{eq:MC-SINRconstraints} can concurrently be satisfied if and only if the spectral radius of $\vect{T}\vect{F}\vect{M}^{-1}+\vect{T}\vect{A}$ is less than unity. Notice that 
      \begin{equation}
      \vect{T}\vect{F}\vect{M}^{-1}+\vect{T}\vect{A}\succeq
      \vect{T}\vect{A}
      \end{equation} 
      which implies that 
      \begin{equation}
      r\left(\vect{T}\vect{F}\vect{M}^{-1}+\vect{T}\vect{A}\right) \geq
      r\left(\vect{T}\vect{A}\right),
      \end{equation}      
      as all of the matrices are non-negative.
      Although increasing the number of antennas will reduce the spectral radius of $\vect{T}\vect{F}\vect{M}^{-1}+\vect{T}\vect{A}$, even in the asymptotic region where the number of antennas approaches infinity the term due to coherent interference, $\vect{T}\vect{A}$, does not vanish. Hence, it is not possible to make the system feasible.
	  \end{IEEEproof}	
  
	  A crucial difference between the single-cell and multi-cell setups is revealed in Lemma \ref{lem:MC-feasibility}. In the single-cell case, $\vect{T}\vect{A} = \vect{0}$ and it is always possible to make a system feasible by increasing the number of antennas whereas in a multi-cell system this is not the case. This phenomenon is a result of the coherent interference caused by pilot contamination, which does not vanish with increasing $M$ and limits the achievable SINRs. The maximum SINR that can be concurrently achieved by all users is summarized in the following lemma. 
	
	  \begin{lemma} \label{lem:MC-maxminLimit}
	  Consider a multi-cell system with $\bar{M}_1 = \bar{M}_2 \ldots =\bar{M}_L = M$ and let $\alpha^*$ denote the maximum SINR that can be achieved by all users. Then, 
	  \begin{equation} \label{eq:MC-maxminSINR}
	  0<\alpha^* \leq \frac{1}{r(\vect{A})}
	  \end{equation}
	  and equality is satisfied only as $M \rightarrow \infty$.
	  \end{lemma}  
  \begin{IEEEproof}
      It is clear that $\alpha^*$ is lower bounded by $0$. As $M \rightarrow \infty$, the non-coherent interference term $\vect{T}\vect{F}\vect{M}^{-1}$ vanishes and for a feasible system $r\left(\vect{T}\vect{A}\right)$ must be less than unity. This gives the upper bound on the $\alpha^*$ stated in \eqref{eq:MC-maxminSINR}.      
	  \end{IEEEproof}
     An interesting property of the two precoding schemes is revealed by Lemma \ref{lem:MC-maxminLimit}. In the asymptotic region, MRT and ZF give the same max-min SINR solution which agrees with the asymptotic analysis given in \cite[Section 4.4.1]{redBook}.       
  
	  \begin{remark}
	  	For a multi-cell system with orthogonal pilots assigned to each user, there is no coherent interference and it is possible to reach a feasible system for any target value by increasing the number of antennas (assuming $\bar{M}_{\max} = \infty$) in the system similar to single cell case.  
	  \end{remark}
	 	     
	Next, we state the main result for the problem defined in \eqref{pr:PR-3}. 
	 \begin{theorem}\label{th:Th-3-MC}
	 	Consider the problem defined in \eqref{pr:PR-3} and assume that there exists at least a feasible solution for $M_l = M_{\max}$ for all $l$. Then, $\y{M}^* = \y{M}_{\max}$ and 
	 	\begin{equation}\vect{p}^* = \left(\vect{M}_{\max} - \vect{T}\vect{F}-\vect{T}\vect{A}\vect{M}_{\max}\right)^{-1}\bs{v}
	 	\end{equation} 
	 	where $\vect{M}_{\max}$ is the antenna matrix with each diagonal element equal to $\bar{M}_{\max}$.
	 \end{theorem}
    \begin{IEEEproof}
	The results of Lemma \ref{lem:MC-diffMdiffP} suggests that increasing the number of antennas results in a lower transmission power vector and it has been shown that it is impossible to transform a feasible system into an infeasible one by increasing the number of antennas. Hence, using the maximum number of antennas at each BS will give us the solution to the problem defined in \eqref{pr:PR-3}.
	\end{IEEEproof}

    \begin{figure}[t]
	\begin{center}
		\includegraphics[trim=2cm 0.0cm 0.4cm 0.1cm,clip=true,scale = 0.6]{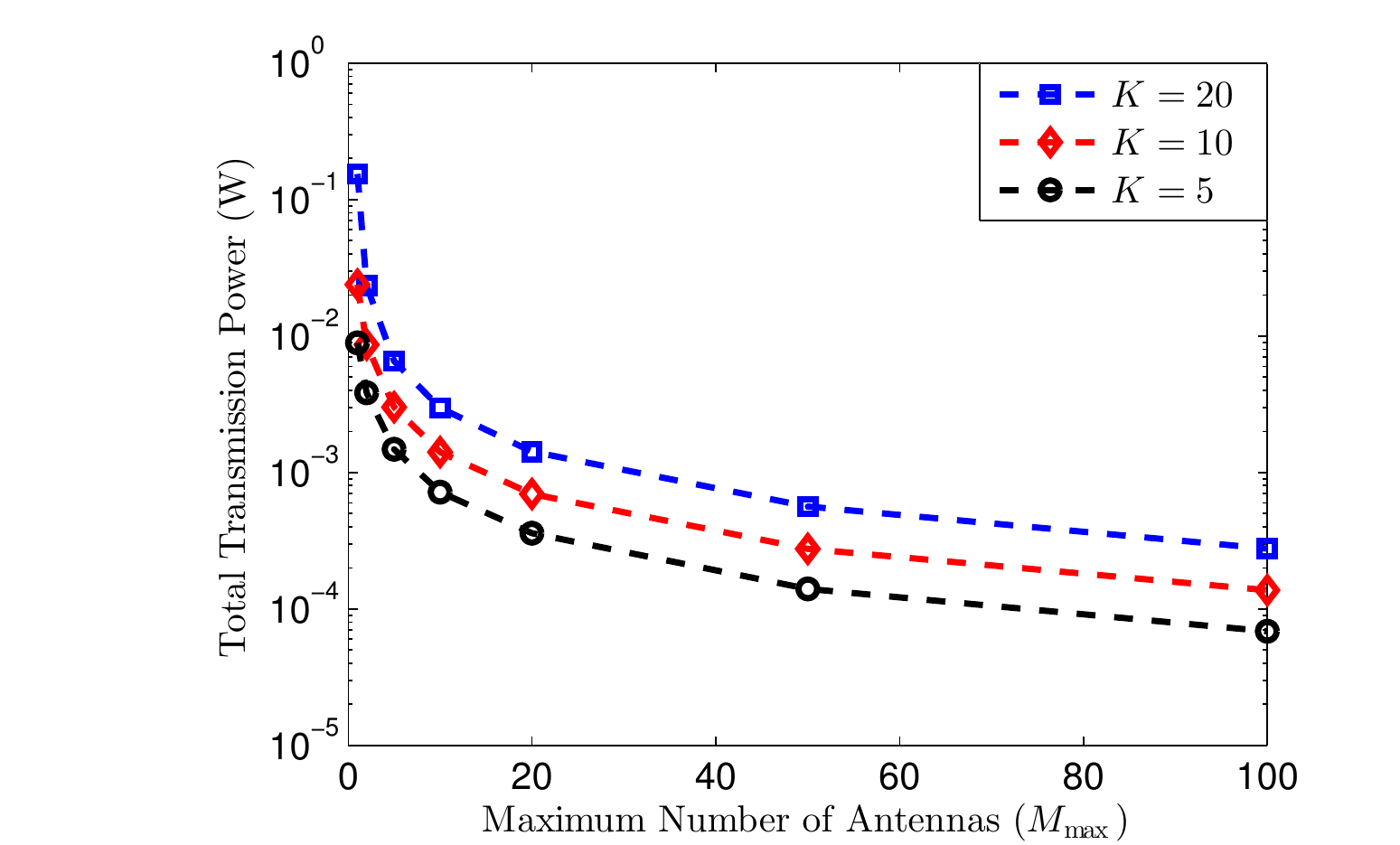}
		\caption{Total transmission power of the optimal solution to \eqref{pr:PR-3} as a function of $M_{\max}$. }
		\label{fig:fig4}
	\end{center}
\end{figure}

    The problem of minimizing only the total transmission power in the system results in a solution which requires to use the maximum number of BS antennas in multi-cell systems, similar to the Theorem \ref{thm:Th-1-SC-Mmax} in the single-cell case. The total transmission power as a function of number of antennas is depicted in Fig.~\ref{fig:fig4} for a setup with MRT. However, the problem becomes more challenging when the cost of utilizing antennas is included in the cost function which is examined next.
    
        The joint optimization problem which also includes a cost for utilizing BS antennas is 
    \begin{equation} \nonumber
    \begin{aligned}
    & \underset{\vect{p}_1,\ldots,\vect{p}_L\succeq 0, \vect{m} \in \mathbb{Z}^+}{\text{minimize}}
    & & c_1\sum_{l = 1}^L\sum_{k = 1}^{K} p_{lk} + c_2\sum_{l = 1}^LM_l \\	 & ~~~~\text{subject to}
    & & \text{SINR}_{lk} \geq \alpha_{lk},~l = 1, \ldots, L,~k = 1,\ldots, K\\
    & & & M_l \leq M_{\max},~l = 1, \ldots, L \\
    & & & \sum_{k = 1}^{K} p_{lk} \leq \rho_{d},~l = 1, \ldots, L.
    \end{aligned}
    \end{equation}
    Consider the following equivalent optimization problem, which can be obtained by dividing the cost function by $c_2$ and utilizing the relative cost of operating each antenna, $c = c_1/c_2$, 
    \begin{equation} \tag{P4}\label{pr:PR-4}
    \begin{aligned}
    & \underset{\vect{p}_1,\ldots,\vect{p}_L\succeq 0, \vect{m} \in \mathbb{Z}^+}{\text{minimize}}
    & & \sum_{l = 1}^L\sum_{k = 1}^{K} p_{lk} + c\sum_{l = 1}^LM_l \\	 & ~~~~\text{subject to}
    & & \text{SINR}_{lk} \geq \alpha_{lk},~l = 1, \ldots, L,~k = 1,\ldots, K\\
    & & & M_l \leq M_{\max},~l = 1, \ldots, L \\
    & & & \sum_{k = 1}^{K} p_{lk} \leq \rho_{d},~l = 1, \ldots, L.
    \end{aligned}
    \end{equation}
    
    It is clear from Theorem \ref{th:Th-3-MC} that increasing number of antennas results in a lower transmission power vector. However, this is not necessarily the optimal solution to \eqref{pr:PR-4} as there is a cost associated with utilizing antennas. Contrary to the single-cell case, it is not possible to obtain a closed-form expression for the solution to \eqref{pr:PR-4}. In principle, we can use \eqref{eq:MC-paretoOptimalSolution} to solve \eqref{pr:PR-4}, as it allows us to find the optimal power vector for a given $\vect{m}$. Considering each $\vect{m}$ in the search space requires examining $(\bar{M}_{\max}-1)^L$ combinations and the corresponding optimal power vector, then choosing the minimum among them 
    would result in finding the optimal solution to \eqref{pr:PR-4}. However, this exhaustive search would not be scalable with $L$ and may not be suitable for practical applications. 
    
    Fortunately, the problem defined by \eqref{pr:PR-4} can be relaxed as a geometric programming (GP) problem for which the optimal solution can be obtained very efficiently and reliably. In order to solve the optimization problem, the required information is on the large scale-coefficients and the SINR constraints. In practice, the GP problem can either be solved by a central unit which conveys the solution to the BSs or each BS can obtain the solution by acquiring the required information from other BSs. A brief introduction to GP is provided in the Appendix \ref{sec:GP} and further details can be found in \cite{boyd2007GP}. 
    
    The objective function of \eqref{pr:PR-4} is a posynomial and it is straightforward to modify all of the constraints except the SINR constraint to obtain a problem at a standard GP form. For the SINR constraint, taking the inverse of the inequality, we obtain 
    \begin{eqnarray}\label{eq:MC-GP_SINR_constraint}
    \frac{1 + \sum\limits_{l' = 1}^L\sum\limits_{k' = 1}^Kp_{l'k'}\beta^{l}_{l'k} + \sum\limits_{l' \in \mathcal{P}_l\backslash \{l\}}M^{}_{l'}\gamma^{l}_{l'k}p^{}_{l'k}}{M_l\gamma^l_{lk}p_{lk}} \leq \frac{1}{\alpha_{lk}},
    \end{eqnarray}
    for MRT. 
    The left-hand side of \eqref{eq:MC-GP_SINR_constraint} is a posynomial. For the case with ZF, the SINR constraints can also be formulated as posynomials in a similar way. Hence, the problem defined by \eqref{pr:PR-4} can be relaxed as a geometric programming problem which can be solved effectively even for many users and large number of antennas. Using the inverse of the SINR to transform the SINR constraints into posynomial constraints have previously been utilized for power control problems to reformulate them as geometric programming problems \cite{boyd2007GP}, but not for the type of problems that we consider here.    
    Similar to the single-cell case the integer constraint on $\vect{m}$ is relaxed and $\vect{m}$ is assumed to be a real-valued vector in the GP problem. In practice, the real-valued vector obtained via solving the GP problem must be converted to an integer-valued vector.

	\section{Numerical Analysis} \label{sec:Numerical}
	In this section, we present the numerical analysis for the multi-cell case which not only verifies the preceding theoretical analysis, but also provides valuable insight for the design of MaMIMO systems. The simulations were performed using Matlab with CVX and the code will be made available online. 
	
\begin{figure}[thb]
		\begin{center}
			\includegraphics[trim=0.8cm 0.1cm 0.2cm 0.1cm,clip=true,scale = 0.25]{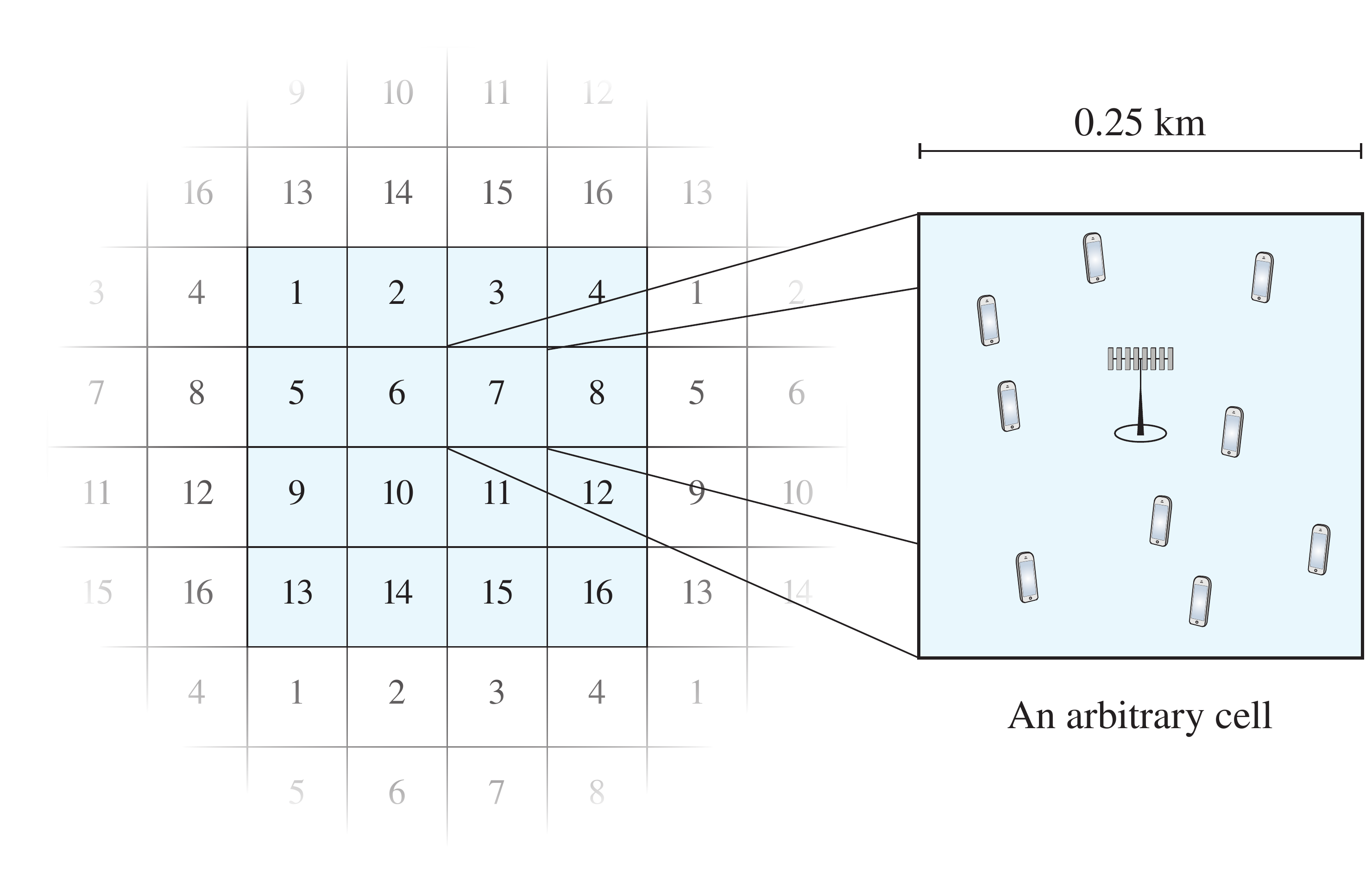}
			\caption{Multi-cell setup with cells distributed on a square grid.}
			\label{fig:MCsetup}
		\end{center}
	\end{figure}

	For the multi-cell simulations, we consider the cellular setup illustrated in Fig. \ref{fig:MCsetup}. There are $L = 16$ cells that are distributed on a $4 \times 4$ square grid with a wrap-around topology. Each cell serves $K$ users that are distributed uniformly in the $250\,$m $\times$ $250\,$m area. There is a minimum distance, $d_{min}$ between users and their serving BS. The rest of the simulation parameters are summarize in Table \ref{tbl:SysParameters}.   
	    
\begin{table}
	\centering
	\caption{Simulation Parameters}
	\label{tbl:SysParameters}
	\begin{tabular}{l|l}
		\hline
		\textbf{System Parameter} & \textbf{Value} \\ \hline
		Path and penetration loss at distance $d$ (km) & 130 + 37.6 $\log_{10}(d)$ \\ 
		Bandwidth ($B_w$)                 & 20 MHz         \\
		Cell Edge Length ($d_{edge}$)          & 250 m          \\
		Minimum Distance ($d_{min}$)          & 15 m           \\ 
		Number of Cells ($L$)                 & 16 \\ 
		Total Noise Power ($B_w \sigma^2$)         & 2$\cdot10^{-13}$ W \\ 
		Maximum DL-Transmission Power ($\rho_{d}$)  & $1$ W        \\
		UL-Transmission Power ($\rho_{ul}$)         & $0.1$ W         \\  
		Relative Pilot Length: $N_p/K$        & 1                \\ 
		Maximum number of BS antennas ($M_{max}$) & 100 \\ \hline
	\end{tabular}
\end{table}

	\begin{figure}[thb]
		\begin{center}
			\includegraphics[trim=0cm 0.1cm 0.2cm 0.1cm,clip=true,scale = 0.6]{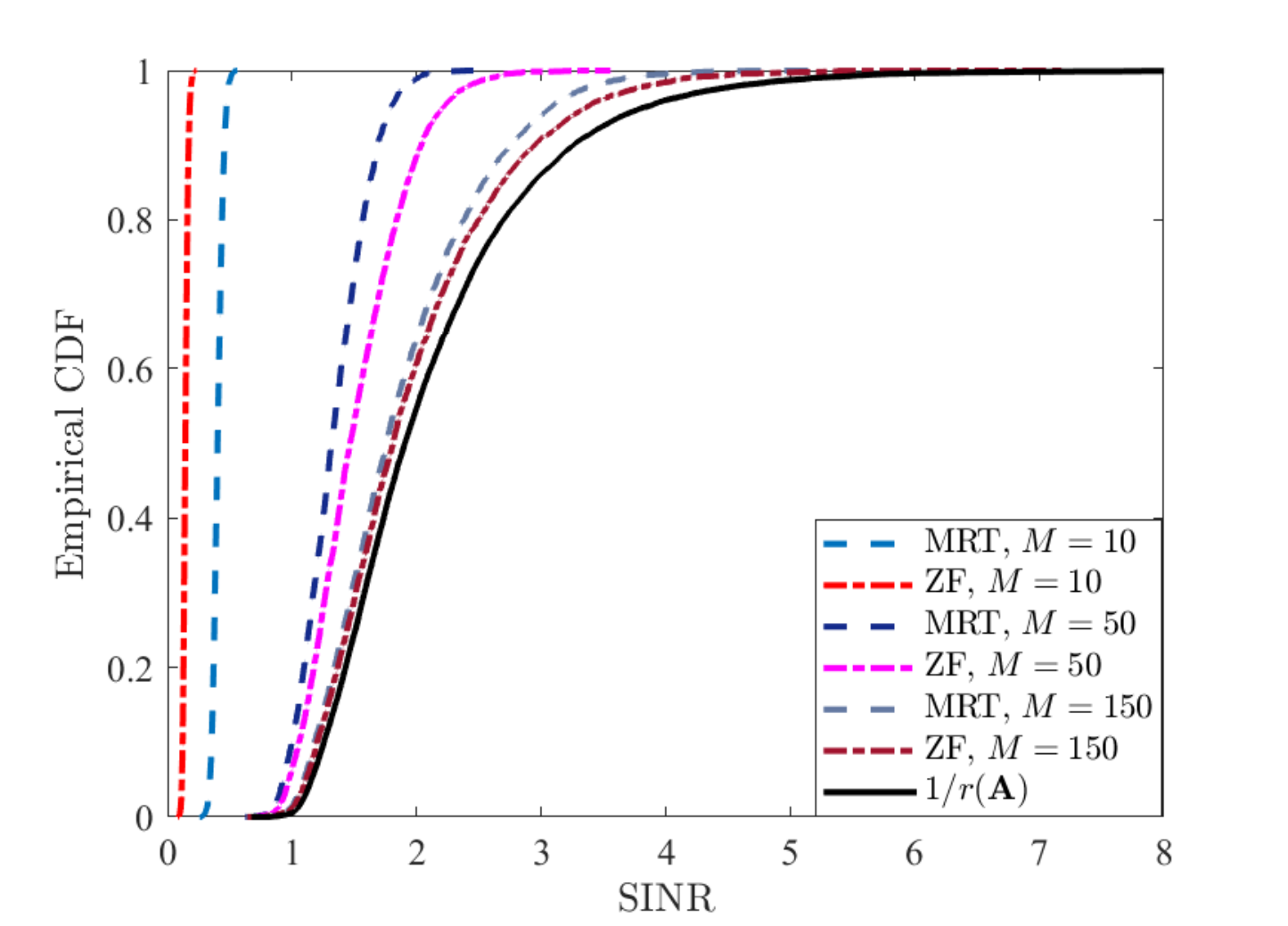}
			\caption{The empirical CDF of max-min SINR for various number of BS antennas.}
			\label{fig:figNew}
		\end{center}
	\end{figure}

   Fig.~\ref{fig:figNew} illustrates the max-min SINR achieved by MRT and ZF for various number of BS antennas. In this particular example, each BS is assumed to have an identical number of antennas. As expected, the achievable max-min SINR increases with the number of BS antennas. However, this increase is limited due to the coherent interference and there is a finite maximum SINR that can be achieved as $M \rightarrow \infty$ as proved by Lemma \ref{lem:MC-maxminLimit}. Fig.~\ref{fig:figNew} also provides an empirical study for the feasibility of a system at a given $M$ and target SINR. For example, with $M = 150$ and target SINR $\alpha = 1.5$, on average $30\%$ of the realizations will be infeasible. This is a consequence of the random user locations and fixed SINR value we selected; in practice, the MAC-layer is responsible for assigning feasible SINR values to the users. Another important point is the performance of the different precoders. MRT provides a better performance when $M$ is low whereas ZF achieves a better performance at higher $M$.

		    \begin{figure}[thb]
		\begin{center}
			\includegraphics[trim=.7cm 0.1cm 0.2cm 0.1cm,clip=true,scale = 0.6]{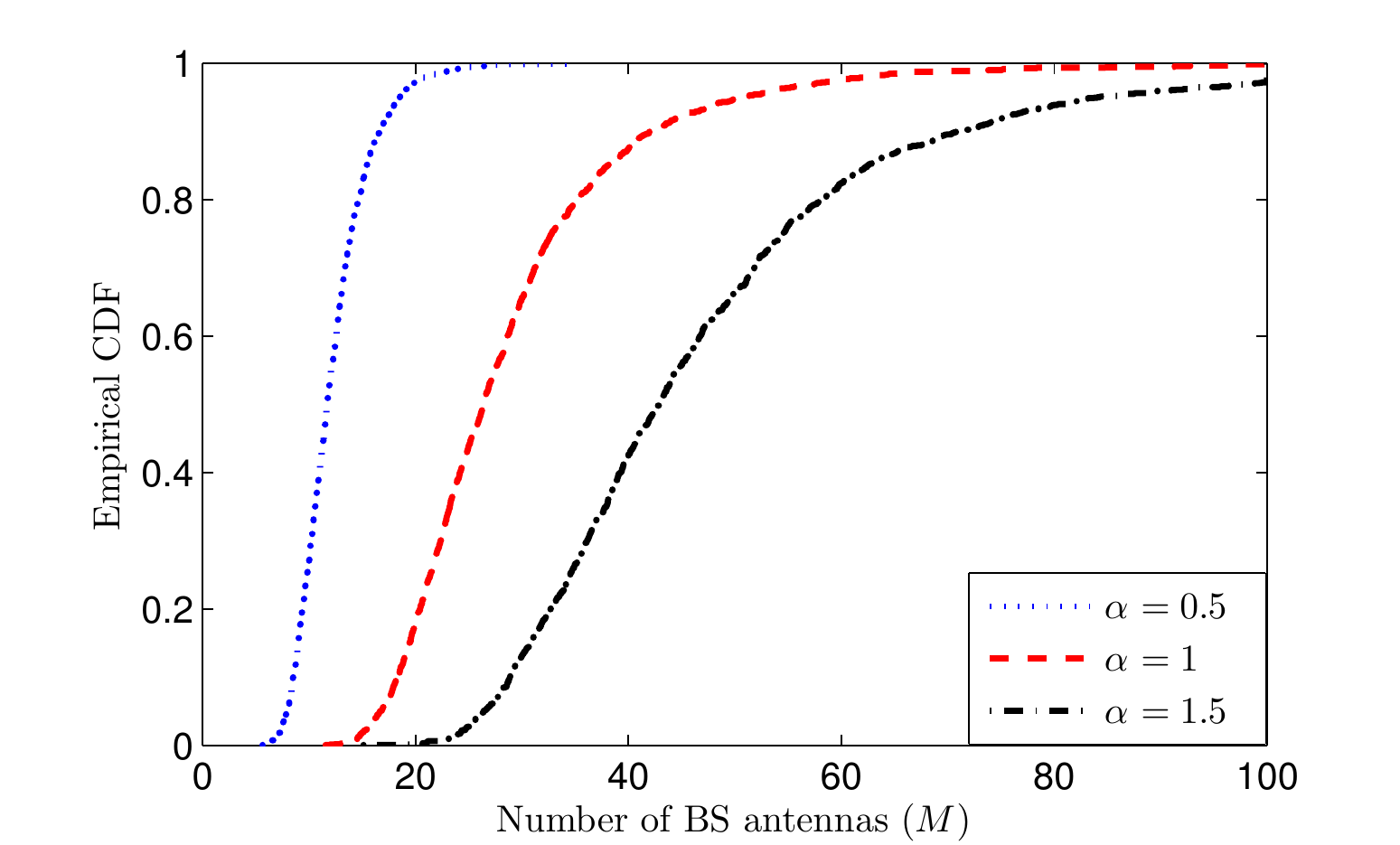}
			\caption{The empirical CDF of the optimal number of BS antennas for \eqref{pr:PR-4} with MRT.}
			\label{fig:fig5}
		\end{center}
	\end{figure}

   Fig.~\ref{fig:fig5} shows the CDF of the optimal number of BS antennas for \eqref{pr:PR-4} with MRT. The simulations are carried out with $K = 8$ and all users have an identical target SINR value, $\alpha$. The optimum solution is obtained by solving the equivalent GP with CVX. As expected, the number of required antennas increases as the target SINR value is increased. Note that some of the trials resulted in an infeasible system for the given target SINR and those trials were discarded.     

\begin{figure}[thb]
	\begin{center}
		\includegraphics[trim=.4cm 0.1cm 0.2cm 0.1cm,clip=true,scale = 0.55]{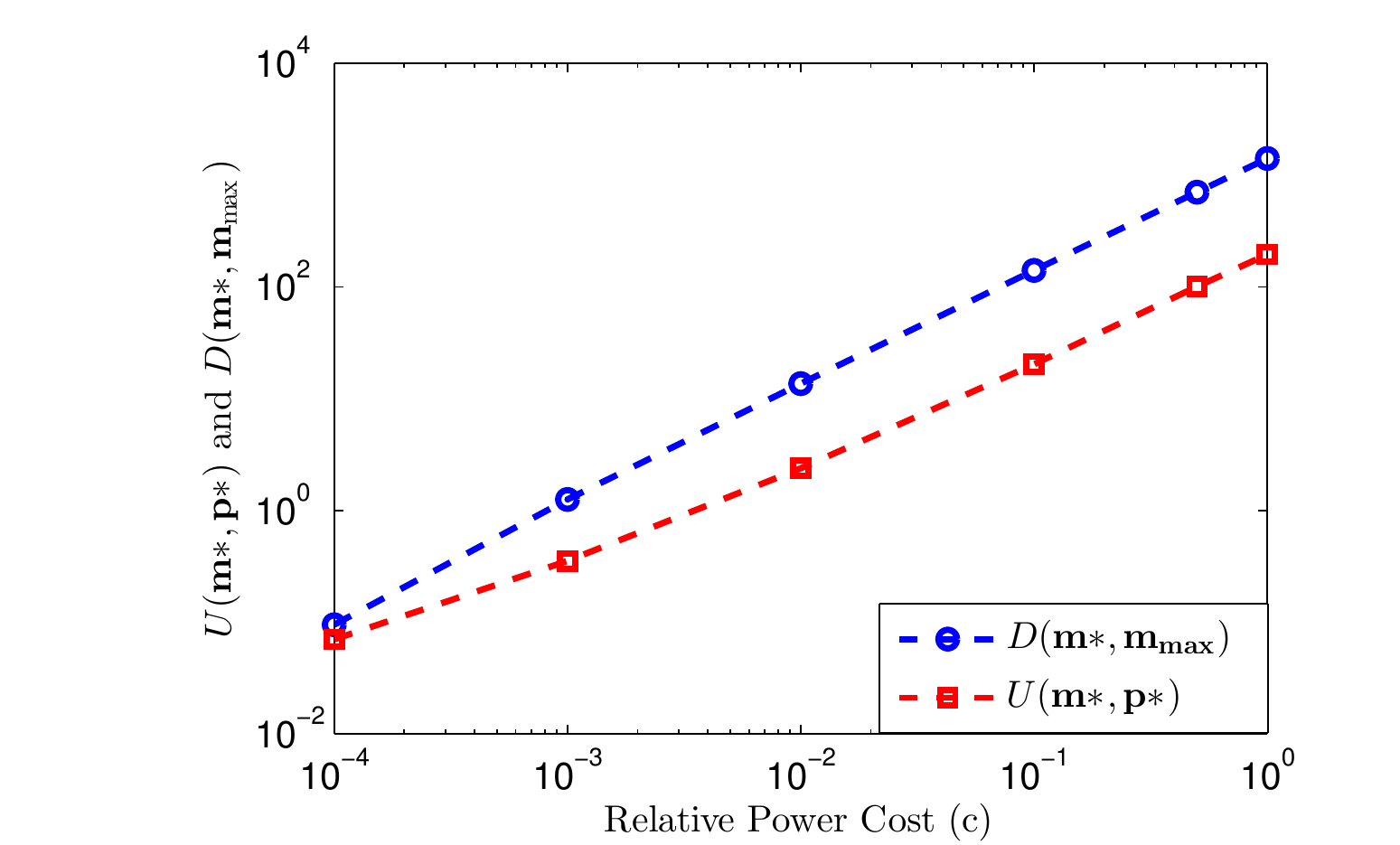}
		\caption{The difference between the cost functions for $\vect{m} = \vect{m}_{\max}$ and the optimal solution along with the cost function of the optimal solution with respect to the relative power cost for \eqref{pr:PR-4}.}
		\label{fig:fig6}
	\end{center}
\end{figure}

Let $D(\vect{m}_1,\vect{m}_2) = |U(\vect{m}_1,\vect{p}_1) - U(\vect{m}_2,\vect{p}_2) |$ be the difference between the two cost function for $\vect{m}_1$ and $\vect{m}_2$ (and the corresponding optimum power vectors). Fig. \ref{fig:fig6} depicts the optimal solution to \eqref{pr:PR-4} obtained via geometric programming along with the difference between the optimum solution and the solution obtained by utilizing the maximum number of antennas, i.e., $U(\vect{m}^*, \vect{p}^*)$ and $D(\vect{m}^*,\vect{m}_{\max})$. There are $K = 8$ users in each cell and MRT precoder is used. The difference between the optimal solution and the solution with maximum number of antennas increases with the relative power cost, as expected. For the case where utilizing more antennas has no cost ($c = 0$), the optimum solution is utilizing the maximum number of antennas at each BS. Assuming a system with $300\,$mW power consumption per antenna and $30\%$ power amplifier efficiency, $c = 0.09$. It is also important to note that based on past trends, the power consumption per antenna scales with a factor of two for each technology generation, whereas the power amplifier efficiency does not benefit from the technological advancements at the same scale which leads to a smaller $c$ value and suggest increased number of BS antennas in future cellular systems \cite{desset2016eeMIMO}, \cite{desset2014PowerConsumption}.  

\begin{figure}[thb]
	\begin{center}
		\includegraphics[trim=.4cm 0.1cm 0.2cm 0.1cm,clip=true,scale = 0.55]{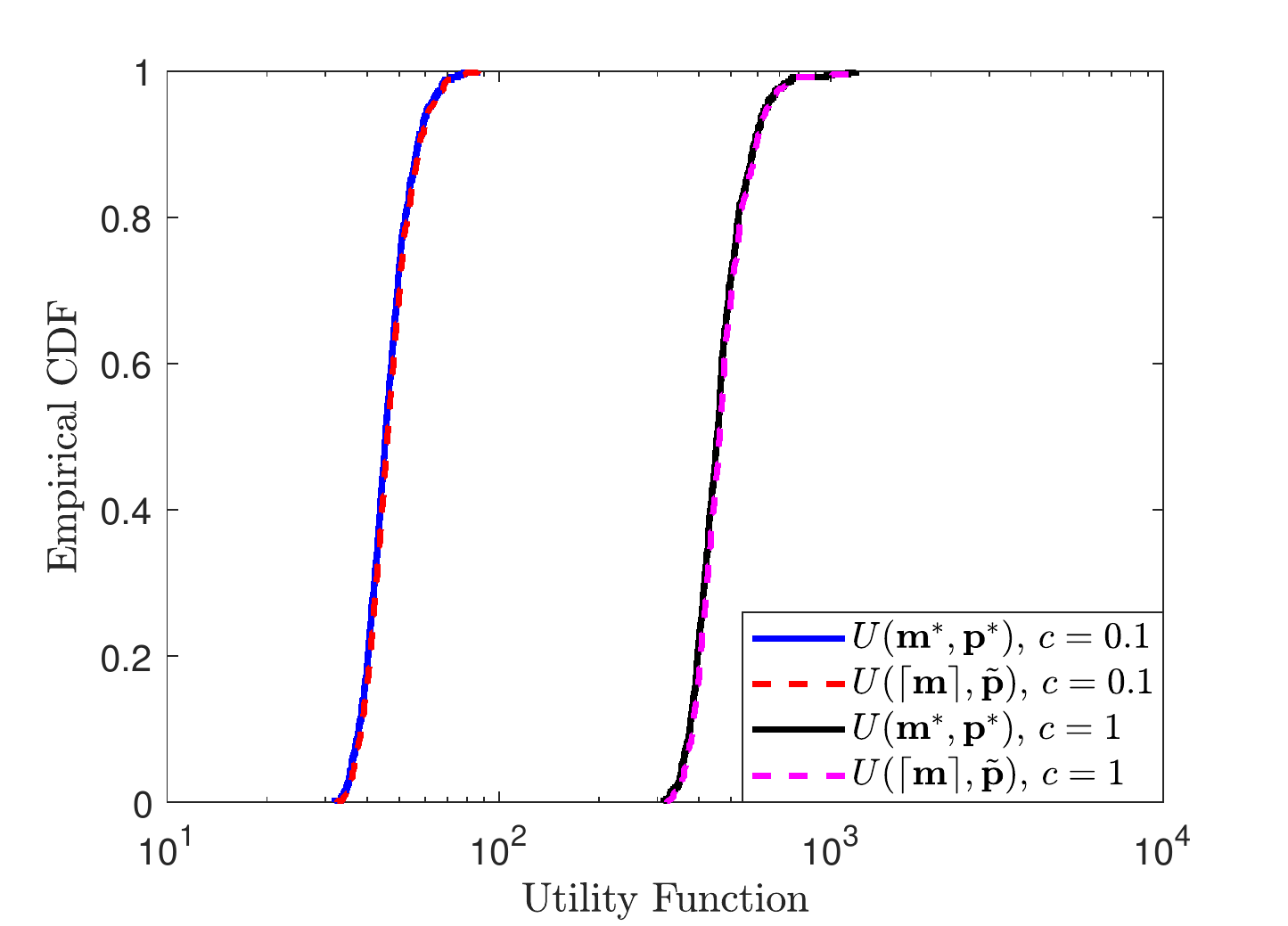}
		\caption{The difference between the cost functions with the integer solution $\vect{m} = \lceil\vect{m}^*\rceil$ and the optimal solution $\vect{m}^*$ for different relative cost values.}
		\label{fig:fig7}
	\end{center}
\end{figure}

In both single-cell and multi-cell cases, the solution is obtained while the integer constraint on the number of antennas is relaxed. However, in a real system the number of active antennas must be integer. In Fig.~\ref{fig:fig7}, the difference between utility functions of the optimal solution obtained via solving the GP problem, $U(\vect{m}^*, \vect{p}^*)$ and the integer solution, $U(\lceil\vect{m}^*\rceil, \tilde{\vect{p}})$, is illustrated. The integer solution, $\lceil\vect{m}^*\rceil$ is obtained via using the ceiling function as this guarantees the feasibility of the solution. Notice that, even though the transmission power $\tilde{\vect{p}}$ for the integer solution is lower than $\vect{p}^*$, the resulting utility function is not. The difference between utility functions is negligible for both $c = 0.1,$ and $c = 1$ which suggests that it is possible to obtain a near optimal solution ceiling the solution obtained by solving the GP problem.   

  \section{Conclusion}\label{sec:Conclusion}
  This work considers the joint optimization of the transmission powers and number of active BS antennas that minimizes the total power consumption in the downlink of a MaMIMO system. We derive a closed-form expression for the optimal solution in single-cell case with MRT and ZF precoding. The optimal operating point depends on the traffic load, energy consumption of each antenna, and the quality of the channel estimates. 
  In the multi-cell case, searching for the optimal solution by employing \eqref{eq:MC-paretoOptimalSolution} for each $\vect{m}$ requires an exhaustive search. Instead, we show that the optimization problem can be reformulated as a geometric programming problem which can be solved efficiently even with a large number of variables. Furthermore, the theoretical analysis reveals that any set of SINR constraints can be satisfied by increasing the number of antennas in the single-cell case. This is not the case in multi-cell case due to pilot contamination. However, even in the existence of coherent interference, it has been shown that the overall system performance in terms of total radiated power can be improved by increasing the number of BS antennas in any cell.
  
   The solutions provided can be used in practice to adapt the system to varying traffic loads, by only turning on the number of antennas required to deliver the requested traffic with minimal power consumption. The uplink analysis could potentially provide similar insights and will be considered  for future work. 
  		
	\appendices
 \section{Matrix Inversion Lemma} \label{sec:MILemma}
The following is a special case of the Matrix inversion lemma (also known as Sherman-Morrison-Woodbury lemma) \cite[pg.~19]{hornMatrixAnalysis}. 
 \begin{lemma}\label{lem:SMV}
       Let $\vect{B}$ be a rank one matrix, then 
       \begin{equation} \label{eq:SMV-BM}
       \left(\vect{I}- \vect{B}\right)^{-1} = \vect{I} + \frac{1}{1 - \text{tr}(\vect{B})}\vect{B}.
       \end{equation} 
 \end{lemma}
	\section{Geometric Programming} \label{sec:GP}
    
    This appendix provides a brief introduction to GP mainly based on \cite{boyd2007GP}. In order to model the optimization problems introduced in this paper, the following definitions are required.
    
    A function is called \emph{monomial} if it is of the form 
    \begin{equation} \label{eq:monomial}
    f(\vect{x}) = \kappa x_1^{a_1}x_2^{a_2}\ldots x_n^{a_n}  
    \end{equation}
    where $\kappa > 0$ and $\vect{x} = [x_1, x_2, \ldots, x_n]^T$ is a vector with real positive components. The exponents $a_1, a_2, \ldots, a_n$ are real valued but not necessarily positive. A function is called \emph{posynomial} if it can be represented as a sum of one or more monomial functions, i.e., 
    \begin{equation}
    f(\vect{x}) = \sum_{k = 1}^K \kappa_k x_1^{a_{1k}}x_2^{a_{2k}}\ldots x_n^{a_{nk}}       
    \end{equation}
    is a posynomial. Note that, addition and multiplication of a posynomial with another one results in a posynomial. Furthermore, division of a posynomial by a monomial gives a posynomial.   
    
    An optimization problem of the form     
    \begin{equation} \label{pr:GP}
    \begin{aligned}
    & {\text{minimize}}
    & & f_0(\vect{x}) \\	 & \text{subject to}
    & & f_i(\vect{x}) \leq 1,~i = 1, \ldots, n,\\
    & & & g_i(\vect{x}) = 1,~i = 1, \ldots, l, \\
   \end{aligned}   
    \end{equation}
    is called a standard form GP if $f_i$ are posynomial functions and $g_i$ are monomial functions of optimization variable $\vect{x}$. If an optimization problem is reduced to the standard form GP, it can be effectively solved by any GP software. In this work, CVX is used \cite{cvx}.   
     
	\ifCLASSOPTIONcaptionsoff
	\newpage
	\fi

\end{document}